\documentclass{jfm}

\usepackage{afterpage}
\usepackage{longtable}
\usepackage[english]{babel}
\usepackage{dcolumn}
\usepackage{bm}
\usepackage{pifont}
\usepackage{amsfonts}
\usepackage{amsmath}
\usepackage{upmath}
\usepackage{amssymb}
\usepackage{wasysym}
\usepackage{amsbsy}
\usepackage{graphicx}
\usepackage{natbib}
\usepackage{psfrag}
\usepackage{lineno}
\usepackage{color}
\usepackage{multirow}
\usepackage{cases}
\usepackage{stackengine}
\usepackage{float}
\usepackage{blkarray}
\usepackage{cancel}
\usepackage{mathrsfs}
\usepackage{mathtools}
\usepackage{empheq}
\usepackage{tikz}
\usetikzlibrary{arrows}
\usetikzlibrary{decorations.pathmorphing}
\usepackage{pgfplots}
\usepackage{psfrag}
\usepackage{subfig}
\usetikzlibrary{quotes,arrows.meta}

\newcommand*\patchAmsMathEnvironmentForLineno[1]{%
  \expandafter\let\csname old#1\expandafter\endcsname\csname #1\endcsname
  \expandafter\let\csname oldend#1\expandafter\endcsname\csname end#1\endcsname
  \renewenvironment{#1}%
     {\linenomath\csname old#1\endcsname}%
     {\csname oldend#1\endcsname\endlinenomath}}%
\newcommand*\patchBothAmsMathEnvironmentsForLineno[1]{%
  \patchAmsMathEnvironmentForLineno{#1}%
  \patchAmsMathEnvironmentForLineno{#1*}}%
\AtBeginDocument{%
\patchBothAmsMathEnvironmentsForLineno{equation}%
\patchBothAmsMathEnvironmentsForLineno{align}%
\patchBothAmsMathEnvironmentsForLineno{flalign}%
\patchBothAmsMathEnvironmentsForLineno{alignat}%
\patchBothAmsMathEnvironmentsForLineno{gather}%
\patchBothAmsMathEnvironmentsForLineno{multline}%
}

\providecommand\bnabla{\boldsymbol{\nabla}}

\providecommand\bcdot{\boldsymbol{\cdot}}
\newcommand{\boldm}[1]{\boldsymbol{#1}}

\title[Start-up flow in shallow deformable microchannels]%
{Start-up flow in shallow deformable microchannels}

\author%
[A. Mart\'inez-Calvo, A. Sevilla, G. G. Peng and H. A. Stone]%
{Alejandro Mart\'inez-Calvo$^1$\thanks{Email address for correspondence: amcalvo@ing.uc3m.es},\ns
Alejandro Sevilla$^1$, \ns Gunnar G. Peng$^2$ \ns
and Howard A. Stone$^3$}

\affiliation{%
$^1$Grupo de Mec\'anica de Fluidos,
Departamento de Ingenier\'ia T\'ermica y de Fluidos,
Universidad Carlos III de Madrid,
Av.~Universidad 30,
28911 Legan\'es (Madrid),
Spain\\[\affilskip]
$^2$Department of Applied Mathematics and Theoretical Physics, University of Cambridge, Wilberforce Road, Cambridge CB3 0WA, UK\\[\affilskip]
$^3$Department of Mechanical and Aerospace Engineering, Princeton University, Princeton, NJ 08544, USA}

\begin{document}
\maketitle

\begin{abstract}
Microfluidic systems are usually fabricated with soft materials that deform due to the fluid stresses. Recent experimental and theoretical studies on the steady flow in shallow deformable microchannels have shown that the flow rate is a nonlinear function of the pressure drop due to the deformation of the upper soft wall. Here, we extend the steady theory of~\citet{Christov2018} by considering the start-up flow from rest, both in pressure-controlled and in flow-rate-controlled configurations. The characteristic scales and relevant parameters governing the transient flow are first identified, followed by the development of an unsteady lubrication theory assuming that the inertia of the fluid is negligible, and that the upper wall can be modeled as an elastic plate under pure bending satisfying the Kirchhoff--Love equation. The model is governed by two non-geometrical dimensionless numbers: a compliance parameter $\beta$, which compares the characteristic displacement of the upper wall with the undeformed channel height, and a parameter $\gamma$ that compares the inertia of the solid with its flexural rigidity. In the limit of negligible solid inertia, $\gamma \to 0$, a quasi-steady model is developed, whereby the fluid pressure satisfies a nonlinear diffusion equation, with $\beta$ as the only parameter, which admits a self-similar solution under pressure-controlled conditions. This simplified lubrication description is validated with coupled three-dimensional numerical simulations of the Navier equations for the elastic solid and the Navier-Stokes equations for the fluid. The agreement is very good when the hypotheses behind the model are satisfied. Unexpectedly, we find fair agreement even in cases where the solid and liquid inertia cannot be neglected.
\end{abstract}


\section{Introduction}
Microfluidic devices allow the manipulation of fluids and objects inside channels whose typical dimensions vary from tens to hundreds of microns. These systems have drastically reduced the working space and the time involved in the applications where they are used. In particular, these devices are routinely used to control multiphase flows, e.g. to generate monodisperse bubbles and droplets, to manipulate immersed soft and hard objects, namely particles, capsules, cells or vesicles, or in diverse applications such as sorting, mixing, drug delivery or mass spectroscopy. The development of microfluidics has facilitated a myriad of applications in a large variety of fields, both in scientific and engineering contexts, e.g. biology~\citep{El2006}, pharmacy and medicine~\citep{Rodriguez2015} or biomedical research~\citep{Sackmann2014}. For more detailed information the reader is referred to the reviews of~\cite{Stone2004},~\cite{Tabeling2005},~\cite{Squires2005},~\cite{Whitesides2006},~\cite{Bruus2008} and~\cite{Anna16}.

The development of soft lithography~\citep{Xia1998} played a crucial role in reducing the manufacturing cost and time of these microfluidic platforms. Soft lithography implies the use of highly flexible materials like polydimethylsiloxane (PDMS), since it is cheap, biocompatible, has a low curing time, and is transparent, which facilitates experimental measurements. These materials are cured on a harder substrate, typically glass, which does not deform appreciably under the characteristic overpressures achieved within the channel, in contrast with the PDMS walls, which may experience a substantial deformation. Hence, the use of soft materials in microfluidics naturally gives rise to coupled fluid-structure interaction (FSI) problems under flow~\citep{Squires2005,Bruus2008,Duprat2015}. In fact, many applications take advantage of the deformation of the compliant walls, whose passive or active control allows the development of soft actuators in lab-on-a-chip devices, e.g. valves, pumps, self-regulating components or flow rectifiers, and has also facilitated the development of soft robotics~\citep{Ilievski2011,Majidi2014,Elbaz2014,Shepherd2011,Rus2015,Polygerinos2017}.

Traditionally, FSI problems have been studied in the context of high-Reynolds-number flows relevant to civil, aeronautical and naval engineering, where aeroelastic and hydroelastic couplings play a crucial role~\citep{Paidoussis2010,Bisplinghoff2013}. In the last few decades, there is a growing attention to FSI problems at small scales due to their ubiquity in nature and in many engineering applications beyond microfluidics. Such is the case of elastocapillarity~\citep{Bico2018}, peeling processes~\citep{Juel2018}, the flow around swimming bacterial colonies~\citep{Lauga2016}, or around vesicles and blood cells in compliant capillaries~\citep{Goldsmith1975,Secomb2002}, where FSI is crucial to understand the underlying physics. A particularly important context where small-scale FSI problems arise is in biological flows~\citep{Fung_3,Fung_1,Fung_2}, e.g. in the pulmonary and respiratory systems~\citep{Grotberg1994,Grotberg2001,Grotberg2004,Heil2011}. Within this context, a large amount of work has been done to understand the dynamics of fluid-filled elastic tubes that may collapse and buckle due to the transmural pressure~\citep{Conrad1969,Shapiro1977,Cancelli1985,Pedley1998,Heil2003}. Although most of these studies deal with finite Reynolds numbers, there are also some works dealing with Stokes flow and lubrication theory in this configuration~\citep{Heil_Pedley1995, Heil1997}. Furthermore, there are still important open problems concerning hemodynamics in vascular networks, involving vascular remodelling, regulation of blood flow, oxygen transport, the fluid pressure distribution or the shear stress exerted at the compliant vessel walls, which are crucial in the development and detection of cardiovascular diseases such as aneurysms or ischemias, or even in tumor angiogenesis~\citep{Goldsmith1975,Pedley1980,Taylor2004,Popel2005,Cassot2006,Lasheras2007,Sforza2009}.

In the present work we focus on the incompressible start-up flow in a shallow microchannel of rectangular cross-section that is filled with a Newtonian liquid, and whose upper soft wall deforms due to the overpressure needed to induce the flow. It is well known that in the low-Reynolds-number flow of liquids inside rigid channels of constant cross section, the flow rate $q$ is proportional to the pressure drop $\Delta p$. The constant of proportionality is usually referred to as the hydraulic resistance, which only depends on the geometry of the cross-section, on the channel length, and on the fluid dynamic viscosity~\citep{Happel2012}. However, several authors have shown through experiments, theory, and simulations that the relationship between $q$ and $\Delta p$ is nonlinear when the deformation of the walls induced by the fluid pressure is not negligible~\citep{Gervais2006,Hardy2009,Seker2009,Cheung2012,Ozsun2013,Raj2016,Raj2017,Christov2018}. Under steady flow conditions, these authors found that, for a given imposed pressure drop, the flow rate is larger than that associated with the corresponding rigid channel. Indeed, higher throughputs can be achieved in deformable microchannels due to a decrease in the hydraulic resistance induced by the wall compliance.

The steady lubrication theory developed by~\cite{Christov2018} assumes that only the upper wall is deformable. Note that, in most microchannels, the bottom wall is rigid, but the lateral ones are made of the same soft material as the upper one, and thus may also deform. Nevertheless, in shallow geometries the deformation of the lateral walls has a negligible effect outside thin lateral elastic boundary layers, as evidenced by the scaling analysis of~\cite{Gervais2006}. In addition, the thickness of the lateral walls is typically much larger than the thickness of the upper wall, in which case the lateral wall deformation is much smaller than the upper one. \citet{Christov2018} described the upper wall as a plate under pure bending modelled with the linear Kirchhoff--Love theory~\citep{Love1888}. Previous studies~\cite[cf.][]{Gervais2006,Hardy2009,Raj2016,Raj2017} considered a global Hookean relation between the fluid pressure and the spanwise average of the upper wall's vertical displacement, where the displacement profile is a quadratic function of the spanwise coordinate, instead of the quartic profile predicted by the Kirchhoff--Love plate theory. Introducing the hypothesised Hookean relation into the standard $q$-$\Delta p$ function, these authors deduced a model with one fitting parameter that absorbs the geometric and material constants. This model was able to explain the trends observed in the experiments of~\cite{Gervais2006} with thick-walled microchannels. However,~\cite{Christov2018} showed that this approximation fails in many configurations, e.g. when the thickness of the top wall is smaller than or comparable to the channel's width. In contrast, the lubrication theory of~\cite{Christov2018} does not have any fitting parameter, and depends only on a compliance parameter $\tilde{\beta}$, that arises naturally from the coupling between the steady fluid flow, described with lubrication theory, and the upper wall displacement, described with the Kirchhoff-Love plate theory. The parameter $\tilde{\beta}$ compares the characteristic displacement of the upper wall with its undeformed height or, equivalently, the characteristic overpressure with the flexural rigidity of the upper wall.~\cite{Christov2018} also made direct comparisons of their theory with the experimental data of~\cite{Ozsun2013}, finding good agreement without fitting parameters. More recently,~\cite{Shidhore2018} have extended these ideas deriving a lubrication model for microchannels with a thicker top wall, whose displacement is modelled with the Mindlin theory accounting for shear stresses. They also performed steady three-dimensional (3D) numerical simulations for both thin and thick upper walls, finding good agreement with their lubrication models, but worse agreement with the experimental data of~\cite{Ozsun2013}.~\cite{Gervais2006} also conducted 3D numerical simulations of the steady flow. However, they did not apply clamped boundary conditions for the top wall displacement at the inlet and outlet of the microchannel, and they did not impose the exact continuity of stresses at the fluid-solid interface.


All the studies mentioned in the previous paragraph focused on steady flow. There are also two previous experimental and theoretical works dealing with unsteady flow in deformable microchannels, namely~\cite{Dendukuri2007} and~\cite{Panda2009}. In particular, both studies considered the stop-flow associated with the relaxation of a top wall initially deflected by the fluid stresses, which induces a squeeze flow towards the inlet and the outlet of the microchannel. To derive their lubrication models, these authors neglected the solid and the liquid inertia, and obtained a nonlinear diffusion equation for the vertical displacement of the upper wall. However, the latter equation is markedly different from the one developed herein, since it is based on a Hookean relation between the vertical displacement of the upper wall and the fluid pressure which is valid in the limit of a very thick wall, but not in the case of a thin plate. In particular,~\cite{Dendukuri2007} determined the relaxation time of the upper wall through scaling arguments, finding good agreement with their own experiments with one fitting parameter, equivalent to that introduced by~\cite{Gervais2006} in the case of steady flow. In the present work, we show that the model developed by~\cite{Dendukuri2007}, although an important step forward, fails to describe the unsteady flow for most geometries and wall materials. Another important contribution of~\cite{Dendukuri2007} was to show that the characteristic time scale only depends on the geometry of the channel and on the fluid and solid properties, but not on the fluid pressure or flow rate, in agreement with the results developed herein. The lubrication model of~\cite{Dendukuri2007} was solved numerically by~\cite{Panda2009}, and compared with stop-flow experiments performed with thick-walled microchannels. The main limitations of the unsteady lubrication theory developed by~\cite{Dendukuri2007} are the same as those of the steady lubrication model of~\cite{Gervais2006}, whose shortcomings were indicated by~\cite{Christov2018}.

An unsteady analysis similar to the one presented herein was performed by~\cite{Elbaz2014} for a thin cylindrical soft shell conveying a viscous fluid. These authors identified the characteristic time scale of the unsteady flow, which is equivalent to the one deduced in the present work. In addition, they developed an unsteady lubrication theory neglecting liquid and solid inertia, and deduced a diffusion equation for the fluid pressure. However, the latter equation is linear, since only small values of the compliance parameter, corresponding to small deformations, were considered by~\cite{Elbaz2014}.


In the present work, we extend the steady theory of~\cite{Christov2018} to account for transient flow, and apply the new framework to the canonical problem of start-up flow from rest. From the theoretical point of view, our main motivation is to provide a framework to tackle FSI problems in laminar internal flows dominated by viscous forces. Indeed, although only the start-up flow is analysed herein for brevity, a similar formalism can be developed to study other transient problems such as stop flows or oscillatory flows. Additional motivation comes from a basic question that, to the best of our knowledge, remains unanswered: what is the start-up time of the flow in a deformable microchannel, and how does it depend on the liquid and solid properties? To that end, we first identify the characteristic hydro-elastic scales and the relevant parameters governing the unsteady flow. Then we develop an unsteady lubrication theory accounting for the solid and liquid inertia, assuming that the upper wall is governed by the Kirchhoff--Love theory in the bending-dominated regime. When the liquid and solid inertia are negligible, we derive a nonlinear diffusion equation for the fluid pressure field. To check the model, we perform 3D direct numerical simulations of the Navier and Navier-Stokes equations for the solid and for the liquid respectively, and compare the results with the quasi-steady lubrication model.

The paper is organised as follows: the flow configuration is described in~\S\ref{sec:flow_configuration}. The mathematical formulation is presented in~\S\ref{sec:3D_numerical_simulations} making use of the Navier and Navier-Stokes equations for the elastic upper wall and the incompressible flow, respectively. In~\S\ref{sec:lubrication} we identify the characteristic scales and the dimensionless parameters governing the flow, and we develop an unsteady lubrication theory for the elasto-hydrodynamic problem, which is further simplified in the quasi-steady limit of negligible liquid and solid inertia. The results are presented in \S\ref{sec:results}, including a comparison between the quasi-steady lubrication model and the 3D numerical simulations. Conclusions are drawn in~\S\ref{sec:conclusions}.

\section{Flow configuration}\label{sec:flow_configuration}

\begin{figure}
\begin{center}
\includegraphics[width=0.65\textwidth]{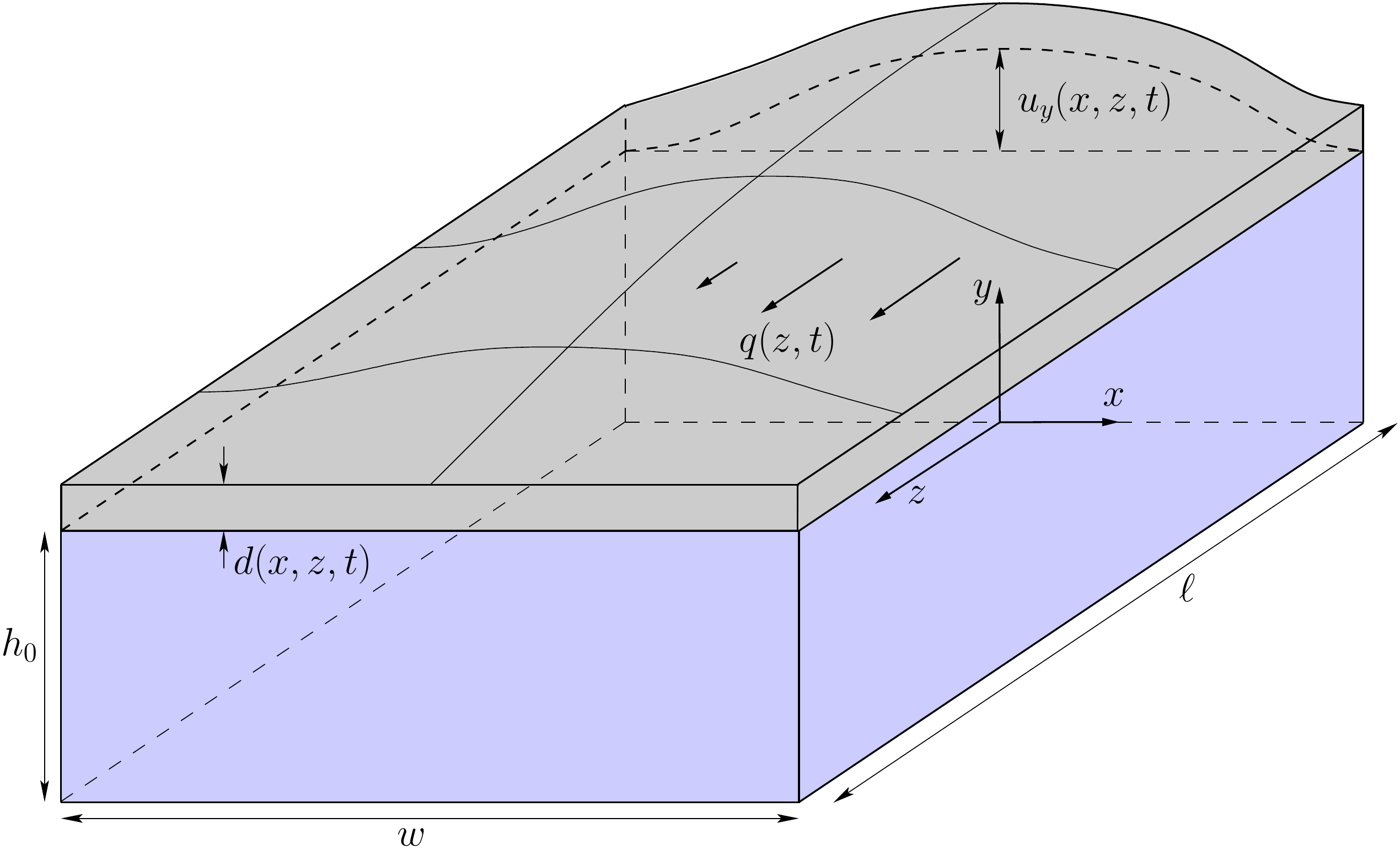}
\caption{(Colour online) Sketch of the flow configuration.\label{fig:fig1}}
\end{center}
\end{figure}

As sketched in figure~\ref{fig:fig1}, we consider the incompressible start-up flow in a channel of length $\ell$, width $w$ and height $h$, where $h \ll w \ll \ell$, initially filled with a Newtonian fluid of density $\rho$ and dynamic viscosity $\mu$. The overpressure needed to convey the fluid deforms the soft walls, which in turn affects the hydraulic resistance of the channel, giving rise to a coupled fluid-structure problem~\citep{Gervais2006,Weibel2007,Hardy2009,Ozsun2013}. Assuming that only the upper wall deforms, here we extend the results of~\cite{Christov2018} to unsteady flow. 

We adopt a Cartesian coordinate system ($x$, $y$, $z$) as shown in figure \ref{fig:fig1}, and use $\boldm{v} = (v_x,v_y,v_z)$ to denote the fluid velocity field and $\boldm{u} = (u_x,u_y,u_z)$ to denote the displacement field of the upper wall, of thickness $d(x,z,t)$. We also use $u_y(x,z,t)$ to denote the vertical displacement of the lower surface of the upper wall, i.e.\ the fluid--solid boundary, so that its position is given by
\begin{equation}\label{eq:interface}
y = h(x,z,t) = h_0 + u_y(x,z,t),
\end{equation}
where $h_0$ is the undeformed height of the channel. The displacement is induced by the fluid pressure $p(x,y,z,t)$, which is measured with respect to the outer atmospheric pressure. The flow rate $q(z,t)$ in the $z$-direction is given by the cross-sectional integral of the axial velocity $v_z$ as
\begin{equation}
q(z,t) = \int_{-w/2}^{w/2} \int_0^{h(x,z,t)}v_z(x,y,z,t) \text{d}y \, \text{d}x.
\end{equation}
For times $t<0$, the fluid is at rest with $p=0$ and $\boldm{v} = \boldm{0}$, and thus the solid remains undeformed, $\boldm{u} = \boldm{0}$. For $t>0$, a start-up flow takes place, either due to an imposed inlet flow rate $q(z = 0,t) = q_0$ (flow-rate-controlled situation) or due to an imposed inlet overpressure $p(z = 0,t) = \Delta p>0$ (pressure-controlled situation). We assume that the outlet pressure is $p(z=\ell,t)=0$. 

To address the start-up flow we have developed two techniques, namely: 1) 3D numerical simulations of the full Navier-Stokes equations for the flow field and the linear Navier equations for the solid deformation field, and 2) a lubrication theory assuming that the upper wall behaves according to the linear Kirchhoff--Love equation for a plate under pure bending~\citep{Love1888}, neglecting changes in its thickness, $d$.

\begin{figure}
\begin{picture}(10,6)
\put(58,-262){\begin{tikzpicture}
\begin{axis}[
hide axis,
scale only axis,
height=0pt,
width=0pt,
colormap/blackwhite,
colorbar/width=2mm,
colorbar horizontal,
point meta min=0,
point meta max=6,
colorbar style={
    width=3.8cm,title={$p$ (kPa)},xticklabel style={align=right},xtick={0,1,2,3,4,5,6}
}]
\addplot [draw=none] coordinates {(0,0)};
\end{axis}
\end{tikzpicture}
}
\put(55,-300.2){
\begin{tikzpicture}
\begin{axis}[
hide axis,
scale only axis,
height=0pt,
width=0pt,
colormap/jet,
colorbar/width=2mm,
colorbar horizontal,
point meta min=0,
point meta max=9.042,
colorbar style={
    width=3.8cm,title={$u_y$ (mm) $\times 10^{-2}$ },at={(0,0)},xticklabel style={align=right},xtick={0,1,2,3,4,5,6,7,8,9}
}]
\addplot [draw=none] coordinates {(0,0)};
\end{axis} \end{tikzpicture}

}
\end{picture}

\begin{tabular}{c}
($a$)\\ 
\hspace{-2cm}\includegraphics[height=0.5\textwidth]{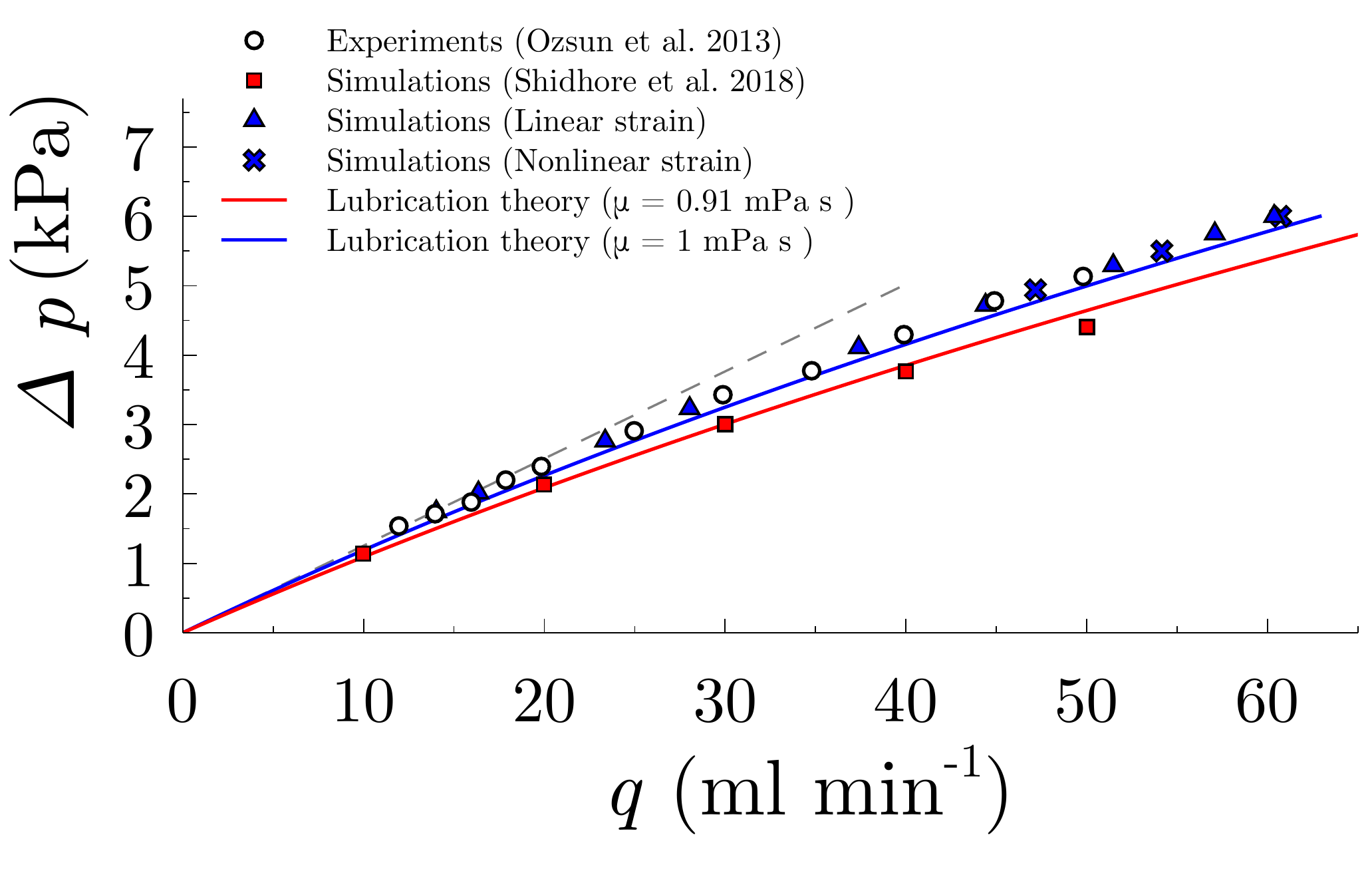} \\
($b$)\\
\hspace{1.5cm} \includegraphics[width=1\textwidth]{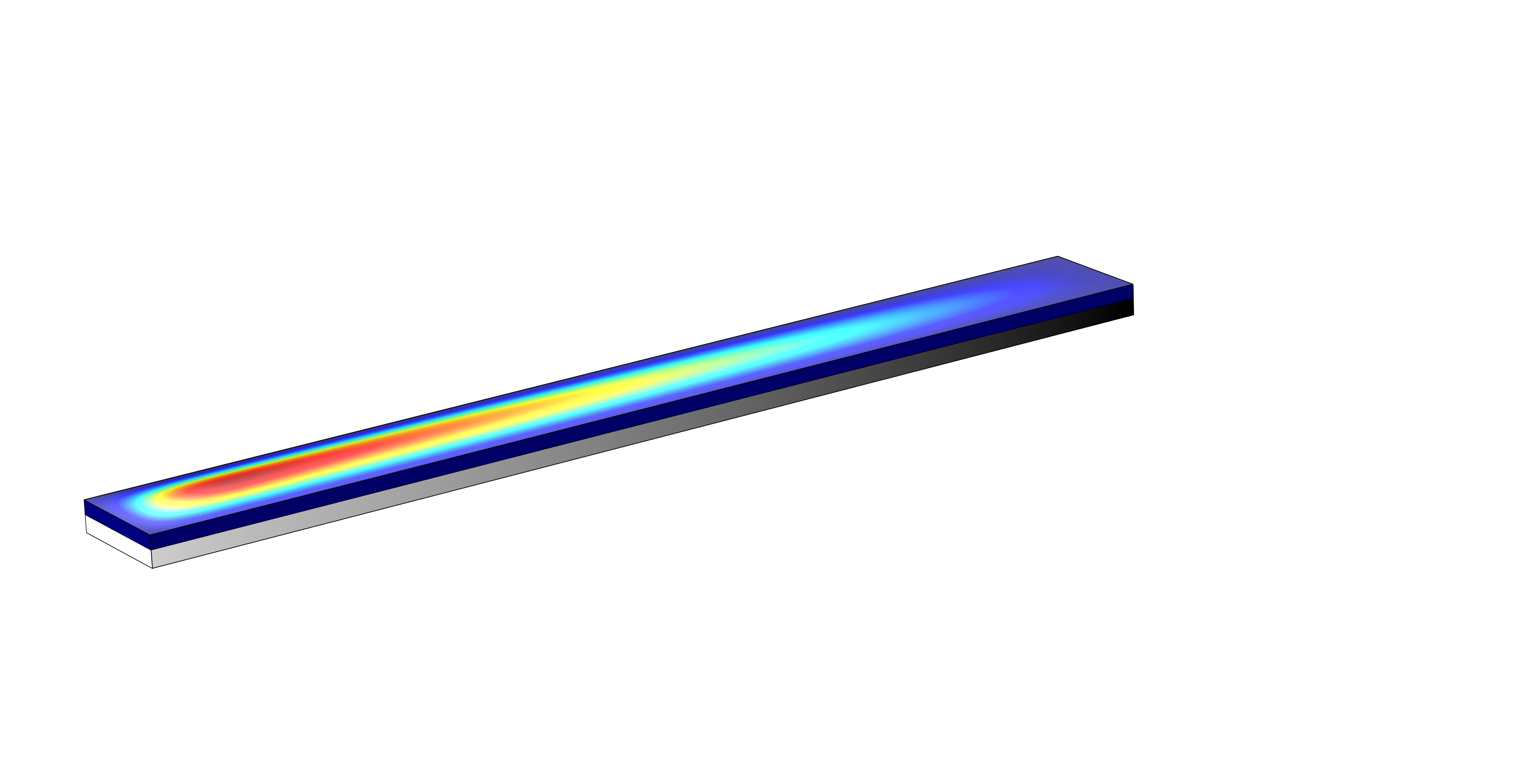}
\vspace{-1.5cm}
\end{tabular}
\caption{\label{fig:fig2} (Colour online) ($a$) Pressure drop as a function of flow rate extracted from the S4 experiments of~\cite{Ozsun2013} (circles), the numerical simulations of~\cite{Shidhore2018} with $\mu = 0.91$ mPa s (red squares), the present numerical simulations considering $\mu = 1$ mPa s with either a linear strain~\eqref{eq:linear_strain} (blue triangles) or a nonlinear strain~\eqref{eq:nonlinear_strain} (blue crosses), and the corresponding lubrication theory of~\cite{Christov2018} for both values of the viscosity (solid lines). The dashed line represents the lubrication solution for rigid channels. ($b$) Steady 3D numerical simulation of the S4 system of~\cite{Ozsun2013} for an imposed inlet pressure of $\Delta p = 6$ kPa, showing the liquid pressure $p$ and the vertical solid displacement $u_y$.}
\end{figure}

\section{Formulation of the problem}\label{sec:3D_numerical_simulations}

Although the main objective of the present study is to develop an unsteady lubrication theory to describe the start-up flow (see~\S\ref{sec:lubrication}), to check its validity we have also performed 3D direct numerical simulations of the Navier equations for a linear elastic upper wall of finite thickness, fully coupled to the Navier-Stokes equations for the incompressible flow of a Newtonian liquid. The fluid velocity field is governed by the continuity and momentum equations,
\begin{equation}\label{eq:continuity_momentum}
\bnabla \bcdot \boldm{v} = 0, \quad \text{and} \quad  \rho (\partial_t \boldm{v} + \boldm{v} \bcdot \bnabla \boldm{v}) = \bnabla \bcdot \mathsfbi{T},
\end{equation}
where $\mathsfbi{T} = -p \mathsfbi{I} + \mu (\bnabla \boldm{v} + \bnabla \boldm{v}^{\text{T}})$ is the fluid stress tensor. The fluid velocity field must satisfy the no-slip conditions at the rigid walls,
\begin{equation}\label{eq:bc1}
\boldm{v} = \boldm{0} \quad \text{at} \quad y = 0, \quad \text{and} \quad x = \pm w/2,
\end{equation}
and atmospheric conditions at the outlet,
\begin{equation}
p = 0 \quad \text{and} \quad \mathsfbi{T} \bcdot \boldm{e}_z = \boldm{0}  \quad \text{at} \quad z = \ell. \label{eq:bc_outlet}
\end{equation}
At the inlet, we either impose a fixed pressure $\Delta p$ and no shear stress,
\begin{equation}
   p = \Delta p \quad \text{and} \quad \mathsfbi{T} \bcdot \boldm{e}_z = \boldm{0} \quad \text{at} \quad z = 0 \quad \text{(pressure-controlled)},\label{eq:ic_bc_1}
\end{equation}
or a given axial flow profile $v_{0}(x,y)$ corresponding to an input flow rate $q = q_0$ (see appendix~\ref{app:numerics}) and no tangential flow,
\begin{equation}
      v_z = v_0(x,y) \quad \text{and} \quad v_x = v_y = 0 \quad \text{at} \quad z =0 \quad \text{(flow-rate-controlled)}. \label{eq:ic_bc_2} \\
\end{equation}
At the contact surface between the liquid and the soft wall, the continuity of velocities must hold,
\begin{equation}\label{eq:bc2}
\boldm{v} = \partial_t \boldm{u} \quad \text{at} \quad y = h(x,z,t).
\end{equation}
The elastic upper wall satisfies the Navier momentum equation,
\begin{equation}\label{eq:navier}
\rho_s \partial^2_t \boldm{u} = \bnabla \bcdot \boldsymbol{\sigma},
\end{equation}
where $\rho_s$ is the upper wall material density, $\boldsymbol{\sigma} = 2 \mu_s \boldsymbol{\varepsilon} + \lambda_s \text{tr} (\boldsymbol{\varepsilon}) \mathsfbi{I}$ is the solid stress tensor, $\boldsymbol{\varepsilon}$ is the strain tensor, and $\mu_s = E/[2(1+\nu)]$ and $\lambda_s = E \nu /[(1+\nu)(1- 2 \nu)]$ are the two Lam\'e constants, expressed in terms of the Young modulus $E$ and Poisson ratio $\nu$. The main results of~\S\ref{sec:results} have been obtained assuming that the strain is linear in the displacement gradients,
\begin{equation}\label{eq:linear_strain}
\boldsymbol{\varepsilon} = \frac{1}{2} \left(\bnabla \boldm{u} + \bnabla \boldm{u}^{\text{T}} \right),
\end{equation}
but we have also used the complete nonlinear expression of the strain tensor,
\begin{equation}\label{eq:nonlinear_strain}
\boldsymbol{\varepsilon} = \frac{1}{2} (\bnabla \boldm{u} + \bnabla \boldm{u}^{\text{T}}+\bnabla \boldm{u}\bcdot \bnabla \boldm{u}^{\text{T}}),
\end{equation}
to check whether stretching of the wall has a significant effect. This nonlinear stretching may be significant even when the deformation gradient $\bnabla\boldm{u}$ and strain $\boldm{\varepsilon}$ are small and the linear stress--strain relationship holds, such as for a thin elastic sheet that is deflected by an amount $u_y$ comparable to or larger than its thickness $d$, which can be modelled using the F\"oppl--von K\'arm\'an equations.

The lateral walls of the solid are clamped, so that $\boldm{u}$ satisfies
\begin{equation}\label{eq:clamped_3D}
\boldm{u} = \boldm{0} \quad \text{at} \quad z = 0, \,\, z = \ell, \quad \text{and} \quad x = \pm w/2.
\end{equation}
At the fluid-solid interface the continuity of stresses must be fulfilled,
\begin{equation}\label{eq:continuity_stress}
(\boldsymbol{\sigma} + \mathsfbi{T}) \bcdot \boldm{n} = \boldm{0} \quad \text{at} \quad y = h(x,z,t), 
\end{equation}
where $\boldm{n}$ is the unit normal vector to the liquid-solid interface. Finally, we impose a stress-free condition at the upper surface of the top wall, 
\begin{equation}\label{eq:stress_free}
\boldm{\sigma} \bcdot \boldm{n}_{\text{ext}} = \boldm{0} \quad \text{at} \quad y = h(x,z,t) + d(x,z,t),
\end{equation}
where $\boldm{n}_{\text{ext}}$ is the unit normal vector to the upper surface of the soft wall. To perform 3D numerical simulations, the complete system of equations~\eqref{eq:continuity_momentum}--\eqref{eq:stress_free} is expressed in weak form and solved with the finite-element software COMSOL Multiphysics employing the Arbitrary Lagrangian-Eulerian (ALE) method. The details of the numerical techniques employed herein are provided in appendix~\ref{app:numerics}.

Figure~\ref{fig:fig2}($a$) shows a first comparison of the two calculational frameworks employed herein under steady-state conditions, namely 3D numerical simulations and lubrication theory~\citep{Christov2018}. In particular, the results show the pressure drop along the channel $\Delta p$ as a function of the flow rate $q$, which is also used as a validation of the numerical method by comparing with the S4 experiment of~\cite{Ozsun2013}. These experiments were performed in a PDMS microchannel with water as a working liquid, and whose relevant physical parameters are given in table~\ref{tab:table1}. Our 3D numerical simulations (blue triangles) are in excellent agreement with the experiments (circles), whereas the lubrication approximation (blue solid line) properly captures the nonlinear trend, but slightly overestimates the flow rate, as already shown by~\cite{Christov2018}. The numerical simulations of~\cite{Shidhore2018} are also shown (red squares), although they were carried out taking a liquid viscosity of $\mu = 0.91$ mPa~s, which corresponds to a room temperature of approximately 24 $^o$C, whereas ours were computed for $\mu = 1$ mPa s, corresponding to a room temperature of 20 $^o$C. The prediction according to lubrication theory for $\mu = 0.91$ mPa s is also shown (red solid line). Hence, it is possible to infer that the experiments of~\cite{Ozsun2013} took place at a room temperature of approximately 20 $^o$C. Furthermore, our numerical simulations were computed taking $\rho = 10^3$ kg m$^{-3}$, and $B \simeq 1.6$ $\mu$J, where $B = E d^3/[12(1-\nu^2)]$ is the bending stiffness of the upper wall. As pointed out by~\cite{Christov2018}, there is uncertainty in the measurements of the bending stiffness $B$, although the curve $q$-$\Delta p$ is less sensitive to changes in $B$ than to changes in the liquid viscosity $\mu$.

Additionally, we have also considered the nonlinear strain~\eqref{eq:nonlinear_strain} in some simulations where $q$ and $\Delta p$ are relatively high (blue crosses). We observe that there is only a small deviation between the linear and nonlinear results, as will be confirmed in~\S\ref{sec:results}, which indicates that the strain is small enough that the linear strain--displacement relationship~\eqref{eq:linear_strain} holds.

A typical 3D simulation is presented in figure~\ref{fig:fig2}($b$), which shows the pressure field $p$ in the fluid domain and the vertical solid displacement field $u_y$ in the solid domain, under the same conditions as the S4 experiment of~\cite{Ozsun2013} for an imposed inlet pressure $\Delta p = 6$ kPa. These results correspond to the triangle at the largest value of $q$ in figure~\ref{fig:fig2}($a$).

\section{Lubrication theory for shallow compliant channels}\label{sec:lubrication}

This section is devoted to the derivation of an unsteady lubrication theory for a slender, shallow and deformable microchannel with rectangular cross-section, using as a starting point the complete set of equations~\eqref{eq:continuity_momentum}--\eqref{eq:stress_free} presented in~\S\ref{sec:3D_numerical_simulations}.

\subsection{Kirchhoff-Love theory for the upper wall deformation}
\label{subsec:kirchhoff}

In developing the lubrication model, the Navier equation~\eqref{eq:navier} will be substituted by an appropriate simplified description, based on plate theory, that takes advantage of the upper wall geometry. In particular, if the maximum displacement of the upper wall is small compared to its thickness $d$, which is constant and smaller than the channel's width $w$, the dynamics of the plate under pure bending can be described with the Kirchhoff--Love equation~\citep{Love1888,Howell2009} with clamped boundary conditions at $x=~\pm w/2$. Therefore, the vertical displacement is now independent of $y$, i.e. $u_y = u_y(x,z)$, and satisfies
\begin{subequations}
\begin{gather}
\rho_s d \, \partial^2_{t} u_y + B \bnabla^4_{xz} u_y = p \quad \text{at} \quad y = h(x,z,t), \label{eq:kirchhoff_love} \\
u_y = 0, \quad \text{and} \quad \partial_z u_y = 0 \quad \text{at} \quad z = 0, \quad \text{and} \quad z = \ell, \label{eq:bc_u1}\\
u_y = 0, \quad \text{and} \quad \partial_x u_y = 0 \quad \text{at} \quad x = \pm w/2, \label{eq:bc_u}
\end{gather}
\end{subequations}
where $\bnabla^4_{xz}$ is the biharmonic operator in the $(x,z)$ plane, and the bending stiffness $B =~E d^3/[12(1-~\nu^2)]$ of the upper wall is assumed constant.

\subsection{Characteristic scales of the unsteady flow}\label{subsec:time_scales}

Here we obtain the characteristic scales and the dimensionless parameters governing the unsteady flow in the coupled elasto-hydrodynamic problem, identifying the conditions under which the set of equations~\eqref{eq:continuity_momentum}--\eqref{eq:bc2} and~\eqref{eq:kirchhoff_love}--\eqref{eq:bc_u} can be approximated by either an unsteady or a quasi-steady lubrication model. First, we set the dominant balances that govern the problem by taking advantage of the geometry of the microchannel and its top wall, which both are slender and narrow, i.e. $h_0 \ll w \ll \ell$, and $d \ll w$. These scales imply the hierarchy $\epsilon \ll \delta \ll 1$, and $\alpha \ll 1$, where $\epsilon=h_0/\ell$, $\delta = h_0/w$ and $\alpha = d/w$ measure the slenderness and the shallowness of the channel, and the narrow geometry of the plate, respectively. The three dominant balances come from, respectively, the standard lubrication force balance in the $z$-direction $\partial_z p \sim \mu \partial_y^2 v_z$, the narrow-geometry plate balance in~\eqref{eq:kirchhoff_love} $B \partial_x^4 u_y \sim p$, and the flux balance $\partial_t u_y \sim v_y \sim \epsilon V_{\ell}$. The latter can be also deduced from the Reynolds equation
\begin{equation}
\int_{-w/2}^{w/2} (\partial_t u_y) \,{\rm{d}}x + \partial_z q = 0, \label{eq:reynolds_eq}
\end{equation} 
where we have used the kinematic condition at the liquid-solid interface, $\boldm{v} \bcdot \boldm{n} = \boldm{n}~\bcdot~\partial_t \boldm{u}$, which yields $\partial_t u_y + v_x \partial_x u_y + v_z \partial_z u_y = v_y$, assuming that the displacement is only in the $y$-direction, $\boldm{u} = (0,u_y,0)$.

Using the balances above, the following relations are obtained:
\begin{equation}
V_{\ell} \sim \frac{h_0^2 p_c}{12 \mu \ell}, \quad u_c \sim \frac{w^4 p_c}{B}, \quad \text{and} \quad t_c \sim \frac{u_c \ell}{h_0 V_{\ell}} \sim \frac{12 \mu \ell^2 w^4}{B h_0^3},
\end{equation}
where $p_c$, $V_{\ell}$, $u_c$ and $t_c$ are the characteristic pressure, axial velocity, top wall displacement, and time scale, respectively. In particular, the characteristic pressure is $p_c \sim \Delta p$ in a pressure-controlled configuration or $p_c \sim 12 \mu \ell q_0/(w h_0^3)$ in a flow-rate-controlled situation, where the factor $12$ is the classical lubrication factor. Note that $t_c$ does not depend on $\Delta p$ or $q_0$, but only on the geometry and on the fluid and solid properties.

The convective acceleration of the fluid can be neglected compared with the viscous force in~\eqref{eq:continuity_momentum} when $O(\rho \boldm{v} \bcdot \bnabla \boldm{v})/O(\mu \bnabla^2 \boldm{v}) \sim \epsilon Re = (h_0/\ell) \rho h_0^3 p_c/(12 \mu^2 \ell) \ll 1$, where $Re$ is the Reynolds number based on the characteristic pressure $p_c$ and on the undeformed height $h_0$. The local acceleration of the fluid in~\eqref{eq:continuity_momentum} is negligible when the viscous diffusion time, $\rho h_0^2/\mu$, is much smaller than the characteristic hydro-elastic time $t_c$, i.e. when $O(\rho \partial_t \boldm{v})/O(\mu \bnabla^2 \boldm{v}) \sim 12 \epsilon Re/\beta = \rho h_0^5 B/(\mu^2 \ell^2 w^4) \ll 1$, which is the so-called Womersley number, and where $\beta =u_c/h_0 = w^4 p_c/(h_0 B)$ is the compliance parameter. Furthermore, the inertia of the top wall can be neglected in~\eqref{eq:kirchhoff_love} when the characteristic time for which the inertia of the solid affects its displacement, $t_s$, is much smaller than the characteristic start-up time involving the deflection of the boundary, $t_c$, i.e. when $O(\rho_s d \, \partial_t^2 u_y)/O(B \bnabla^4_{xz} u_y) \sim~\gamma = (t_s/t_c)^2 =~\rho_s d h_0^6 B/(\mu^2 \ell^4 w^4) \ll 1$. Note that, like $t_c$, the dimensionless numbers $\epsilon Re /\beta$ and $\gamma$ do not depend on $\Delta p$ or $q_0$, but only on the geometry of the channel and on the solid and fluid properties.

\begin{table}
\centering
\begin{tabular}{c c c c c c c c c}
$y$, $x$, $z$ & $v_y$, $v_x$, $v_z$~\eqref{eq:continuity_dimless} & Equation~\eqref{eq:momentumZ_dimless} & Equation~\eqref{eq:displacement_dimless} & Equation~\eqref{eq:reynolds_dimless} \\
&\\
$h_0 \ll w \ll \ell$, $d \ll w$ & $\dfrac{h_0}{\ell} V_{\ell} \ll \dfrac{w}{\ell} V_{\ell} \ll V_{\ell}$ & $V_{\ell} \sim\dfrac{h_0^2 p_c}{12 \mu \ell}$ & $u_c\sim\dfrac{w^4 p_c}{B}$ & $t_c\sim\dfrac{\mu w^4 \ell^2}{B h_0^3}$ \\ \hline
\end{tabular}
\setlength{\tabcolsep}{6pt}
\begin{tabular}{ccccc}	
$\epsilon \ll \delta \ll 1$, $\alpha \ll 1$ & $\dfrac{O(u_y)}{h_0}$ & $\dfrac{O(\rho \boldm{v} \bcdot \bnabla \boldm{v})}{O(\mu \bnabla^2 \boldm{v})}$ & $\dfrac{O(\rho \partial_t \boldm{v})}{O(\mu \bnabla^2 \boldm{v})} $ & $\dfrac{O(\rho_s d \, \partial_t^2 u_y)}{O(B \bnabla_{xz}^4 u_y)} $\\
&\\
$\epsilon=\dfrac{h_0}{\ell}, \delta=\dfrac{h_0}{w}$, $\alpha = \dfrac{d}{w}$ & $\tilde{\beta} = \dfrac{w^4 p_c}{240 h_0 B}$ & $\dfrac{\epsilon Re}{12} = \dfrac{\rho h_0^4 p_c}{144 \mu^2 \ell^2}$ & $\dfrac{\epsilon Re}{4 \tilde{\beta}} = \dfrac{5 \rho h_0^5 B}{\mu^2 \ell^2 w^4}$ & $\gamma = \dfrac{\rho_s d h_0^6 B}{\mu^2 \ell^4 w^4}$
\end{tabular}
\caption{Characteristic scales used in the derivation of the unsteady lubrication model of~\S\ref{subsec:nondim} (upper row), and the dimensionless parameters that govern the elasto-hydrodynamic problem, deduced in~\S\ref{subsec:time_scales}\label{tab:table2} (bottom row).}
\end{table}




As shown below, a more precise estimate provides $t_c \sim \mu \ell^2 w^4/(60 B h_0^3)$, which leads to a modified parameter $\tilde{\gamma} = 60^2 \gamma$. It also proves convenient to define a modified compliance parameter $\tilde{\beta}=\beta/240$, which is ten times smaller than the one used by~\cite{Christov2018}. As will be shown in~\S\ref{subsec:nondim}, a more accurate estimate of the local and convective inertia is $\epsilon Re/(4 \tilde{\beta})$, and $\epsilon Re/12$, respectively. Taking typical values from the experimental data of~\cite{Ozsun2013}, who used water as the working liquid, $\tilde{\beta} \sim 1$, $\epsilon Re \sim 1$ and $\tilde{\gamma} \sim 1$. However, it is important to note that $\epsilon Re/(4\tilde{\beta}) \propto \mu^{-2}$ and $\tilde{\gamma}\propto \mu^{-2}$, so that both dimensionless parameters rapidly become small as the liquid viscosity increases. Therefore, the flow is governed by six dimensionless parameters, namely $\epsilon$, $\delta$, $\alpha$, $\tilde{\beta}$, $\epsilon Re$, and $\tilde{\gamma}$,
\begin{equation}\label{eq:parameters}
\epsilon = \frac{h_0}{\ell}, \quad \delta = \frac{h_0}{w}, \quad \alpha = \frac{d}{w},\quad \tilde{\beta} = \frac{w^4 p_c}{240 h_0 B}, \quad \epsilon Re = \frac{\rho h_0^4 p_c}{12 \mu^2 \ell^2}, \quad \tilde{\gamma} = \frac{60^2 \rho_s d h_0^6 B}{\mu^2 \ell^2 w^4},
\end{equation}
where the only non-geometrical parameters are $\tilde{\beta}$, $Re$ and $\tilde{\gamma}$. Table~\ref{tab:table2} summarises the analysis on the characteristic scales presented herein, and the dimensionless parameters governing the problem. 

Characteristic times equivalent to those deduced above have been obtained previously. For instance, in the same configuration as ours,~\cite{Dendukuri2007} deduced similar scalings under the assumption that the spanwise average of the upper wall's vertical displacement is linearly proportional to the fluid pressure. However, as shown below, such an assumption leads to a free parameter and cannot describe most channel geometries. Furthermore, following the latter procedure,~\cite{Tabeling2005} deduced the characteristic time scale of a deformable cylindrical chamber pumping fluid into a much smaller microfluidic channel, which gives rise to the \textit{bottleneck effect}, where, due to the deformation of the reservoir, the start-up time increases from minutes to hours. In contrast,~\cite{Elbaz2014} obtained the characteristic start-up time without adjustable parameters, following a procedure similar to that developed in the previous paragraph, but in the case of a cylindrical elastic tube conveying a viscous fluid.

\subsection{Non-dimensional formulation}\label{subsec:nondim}

We define the following dimensionless variables 
\begin{subequations}\label{eq:dimless}
\begin{gather}
X = \frac{x}{w}, \quad Y = \frac{y}{h_0}, \quad Z = \frac{z}{\ell}, \quad T = \frac{60 \, h_0^3 B}{\mu w^4 \ell^2}\,t, \quad H = \frac{h}{h_0}, \quad U_Y = \frac{B}{w^4 p_c} \, u_y,\\
V_X = \frac{\delta}{\epsilon} \frac{12 \mu \ell}{h_0^2 p_c}\,v_x, \quad V_Y = \frac{1}{\epsilon} \frac{12 \mu \ell}{h_0^2 p_c} \, v_y, \quad V_Z = \frac{12 \mu \ell}{h_0^2 p_c} v_z, \quad P = \frac{p}{p_c}, \quad Q = \frac{12 \mu \ell}{w h_0^3 p_c}q,
\end{gather}
\end{subequations}
where $p_c = \Delta p$, in a pressure-controlled situation, and $p_c = 12 \mu \ell q_0/(w h_0^3)$ in a flow-rate-controlled configuration. Introducing these variables into~\eqref{eq:continuity_momentum},~\eqref{eq:kirchhoff_love}, and~\eqref{eq:reynolds_eq} provides the following dimensionless equations
\begin{subequations}
\begin{gather}
\bnabla \bcdot \boldm{V} = 0, \label{eq:continuity_dimless} \\
12 \left[ \frac{\epsilon^2}{\delta^2}\left(\frac{\epsilon Re}{4  \tilde{\beta}} \, \partial_T V_X + \frac{\epsilon Re}{12} \, \boldm{V} \bcdot \bnabla V_X \right)+ \partial_X P \right] = \epsilon^2 \partial_X^2 V_X + \frac{\epsilon^2}{\delta^2}\partial_Y^2 V_X + \frac{\epsilon^4}{\delta^2} \partial_Z^2 V_X,\label{eq:momentumX_dimless}\\
12 
\left[ \epsilon^2 \left( \frac{\epsilon Re}{4\tilde{\beta}}  \, \partial_T V_Y + \frac{\epsilon Re}{12} \, \boldm{V} \bcdot \bnabla V_Y \right) + \partial_Y P \right] =  \epsilon^2 \delta^2 \partial_X^2 V_Y + \epsilon^2 \partial_Y^2 V_Y + \epsilon^4 \partial_Z^2 V_Y ,\label{eq:momentumY_dimless}\\
12 \left(\frac{\epsilon Re}{4 \tilde{\beta}} \, \partial_T V_Z + \frac{\epsilon Re}{12} \, \boldm{V} \bcdot \bnabla V_Z +\partial_Z P \right) = \delta^2 \partial_X^2 V_Z + \partial_Y^2 V_Z + \epsilon^2 \partial_Z^2 V_Z, \label{eq:momentumZ_dimless}\\
\tilde{\gamma} \partial_T^2 U_Y + \partial_X^4 U_Y + \frac{\epsilon^2}{\delta^2} 2 \partial_{X}^2 \partial_{Z}^2 U_Y + \frac{\epsilon^4}{\delta^4} \partial_Z^4 U_Y = P, \label{eq:displacement_dimless} \\
720 \int_{-1/2}^{1/2}  (\partial_T U_Y) {\rm{d}}X + \partial_Z Q = 0, \label{eq:reynolds_dimless} \\
Q(Z,T)= \int_{-1/2}^{1/2} \int_0^{1+\beta U_Y(X,Z,T)} V_Z\, {\rm{d}}Y \,{\rm{d}}X, \label{eq:flowrate_dimensionless}
\end{gather}
\end{subequations}
together with the dimensionless version of the boundary conditions, which are omitted here for simplicity. Here $\boldm{V}=(V_X,V_Y,V_Z)$ is the dimensionless velocity vector. 

The elasto-hydrodynamic timescale $t_c$ is the characteristic time for the channel to transition from the undeformed state to the deformed steady state. For a given channel geometry and working fluid, as the rigidity of the wall increases ($B \to \infty$ and hence $\beta \to 0$) the timescale $t_c$ decreases, reflecting the fact that less time is required to inflate the channel to the less deformed final steady state ($u_c \to 0$). For a perfectly rigid channel ($\beta = 0$), the transition occurs instantaneously (in the absence of fluid inertia). Since $t$ is non-dimensionalised using $t_c$~\eqref{eq:dimless}, in the limit $\beta \to 0$ the non-dimensional equations obtained below remain regular and tend towards a limiting solution, which we investigate in appendix~\ref{app:asymp}.

\subsection{Leading-order lubrication model}\label{subsec:lubmod}

Assuming $\epsilon \ll \delta \ll 1$, $\epsilon Re\ll 1$ and $\tilde{\beta} \gg \epsilon Re$, the $X$- and $Y$-momentum equation~\eqref{eq:momentumX_dimless} and~\eqref{eq:momentumY_dimless} yield $\partial_X P = \partial_Y P =0$, so that the pressure field is only a function of the axial coordinate $Z$ and time $T$, i.e. $P = P(Z,T)$. Therefore, at leading order, the set of equations~\eqref{eq:continuity_dimless}--\eqref{eq:flowrate_dimensionless}, yields the following system of nonlinear differential equations,
\begin{subequations}
\begin{gather}
\partial_Z P = \partial^2_Y V_Z, \label{eq:momentumz_dimless} \\
\tilde{\gamma} \, \partial^2_T U_Y + \partial^4_X U_Y = P, \label{eq:kirchhoff_love_dimless} \\
720 \int_{-1/2}^{1/2} (\partial_T U_Y) {\rm{d}}X + \partial_Z Q = 0, \label{eq:reynolds_eq_dimless} \\
Q(Z,T)= \int_{-1/2}^{1/2} \int_0^{1+\beta U_Y(X,Z,T)} V_Z\, {\rm{d}}Y \,{\rm{d}}X. \label{eq:flowrate_adim}
\end{gather}
\end{subequations}
Note that the leading-order lubrication equations~\eqref{eq:momentumz_dimless}--\eqref{eq:flowrate_adim} cannot fulfill all of the boundary conditions of the full set of equations. In particular, the no-slip boundary condition~\eqref{eq:bc1} is not satisfied at the side walls $X=\pm 1/2$, and from the kinematic condition~\eqref{eq:bc2}, only the no-slip condition $V_Z = 0$ remains to be imposed at $Y=H(X,Z,T)$. In addition, the clamped conditions~\eqref{eq:bc_u1} cannot be imposed on the upper wall. Therefore, the side-wall boundary conditions for $V_Z$, and the inlet and outlet boundary conditions for $U_Y$, lead to corrections of the order of $O(\delta)\ll 1$ and $O(\epsilon/\delta)\ll 1$ respectively, as discussed by~\cite{Christov2018}. Hence, equations~\eqref{eq:momentumz_dimless}--\eqref{eq:flowrate_adim} must be complemented with the remaining boundary and initial conditions,
\begin{subequations}
\begin{gather}
V_Z = 0 \quad \text{at} \quad Y = 0, \quad \text{and} \quad Y = H(X,Z,T) = 1 + \beta U_Y(X,Z,T) , \label{eq:bc_Vz_dimless}\\
U_Y = 0, \quad \text{and} \quad \partial_X U_Y = 0 \quad \text{at} \quad X = \pm 1/2, \label{eq:bc_UY_dimless}\\
\begin{rcases} 
      Q = 1, & \text{flow-rate-controlled} \\
      P = 1, & \text{pressure-controlled}
\end{rcases} \quad \text{at} \quad Z = 0, \label{eq:ic_bc_dimless} \\
P = 0 \quad \text{at} \quad Z = 1. \label{eq:bc_outlet_dimless}
\end{gather}
\end{subequations}
In the present work, instead of tackling the complex set of equations~\eqref{eq:momentumz_dimless}--\eqref{eq:bc_outlet_dimless}, we will just consider the case of negligible solid inertia, corresponding to the limit $\tilde{\gamma}\to 0$.

\subsection{The limit $\tilde{\gamma} \to 0$}\label{subsec:quasi_steady}

In the limit $\tilde{\gamma} \to 0$, equations~\eqref{eq:momentumz_dimless}--\eqref{eq:bc_outlet_dimless} provide, at leading order, the quasi-steady description
\begin{subequations}
\begin{gather}
V_Z(X,Y,Z) = -6 Y[1 + \beta U_Y(X,Z,T)-Y]\partial_Z P, \label{eq:axial_velocity} \\
U_Y(X,Z,T) = \frac{P(Z,T)}{24}\left(X^2-\frac{1}{4}\right)^2, \label{eq:displacement} \\
Q(Z,T) = -\partial_Z P \int_{-1/2}^{1/2} [1+\beta U_Y(X,Z,T)]^3 {\rm{d}}X = - F(\tilde{\beta}P) \partial_Z P, \label{eq:flow_rate}
\end{gather}
\end{subequations}
where
\begin{equation}\label{eq:F1}
F(x)=1 + x +\frac{10}{21} x^2+\frac{250}{3003} x^3, \quad \text{and} \quad \tilde{\beta} = \frac{\beta}{240}.
\end{equation}
The Reynolds equation~\eqref{eq:reynolds_eq_dimless} yields, using~\eqref{eq:displacement} and~\eqref{eq:flow_rate}, a nonlinear diffusion equation for $P(Z,T)$,
\begin{equation}\label{eq:P_eq}
\partial_{T} P = \partial_Z [ F(\tilde{\beta}P)\partial_Z P],
\end{equation}
subject to the boundary and initial conditions
\begin{subequations}
\begin{gather}
\begin{rcases} 
      P = 1, & \text{pressure-controlled} \\
      -F(\tilde{\beta} P)\partial_Z P = 1, & \text{flow-rate-controlled}
\end{rcases} \quad \text{at} \quad Z = 0, \label{eq:a} \\
P = 0 \quad \text{at} \quad Z = 1, \label{eq:b}\\
P = 0 \quad \text{at} \quad T = 0, \quad \text{and} \quad 0<Z<1. \label{eq:c}
\end{gather}
\end{subequations}
Note that if the transient is flow-rate-controlled, the dimensionless inlet pressure is a function of time, $P_0(T) = P(Z=0,T)$ in order to impose a constant inlet flow rate $Q_0 = Q(Z=0,T) = 1$, whose expression can be obtained from~\eqref{eq:flow_rate}.

\subsection{Self-similar solutions}\label{subsec:self_similar}

\begin{figure}
\begin{tabular}{cc}
($a$) & ($b$)\\
 \includegraphics[width=0.5\textwidth]{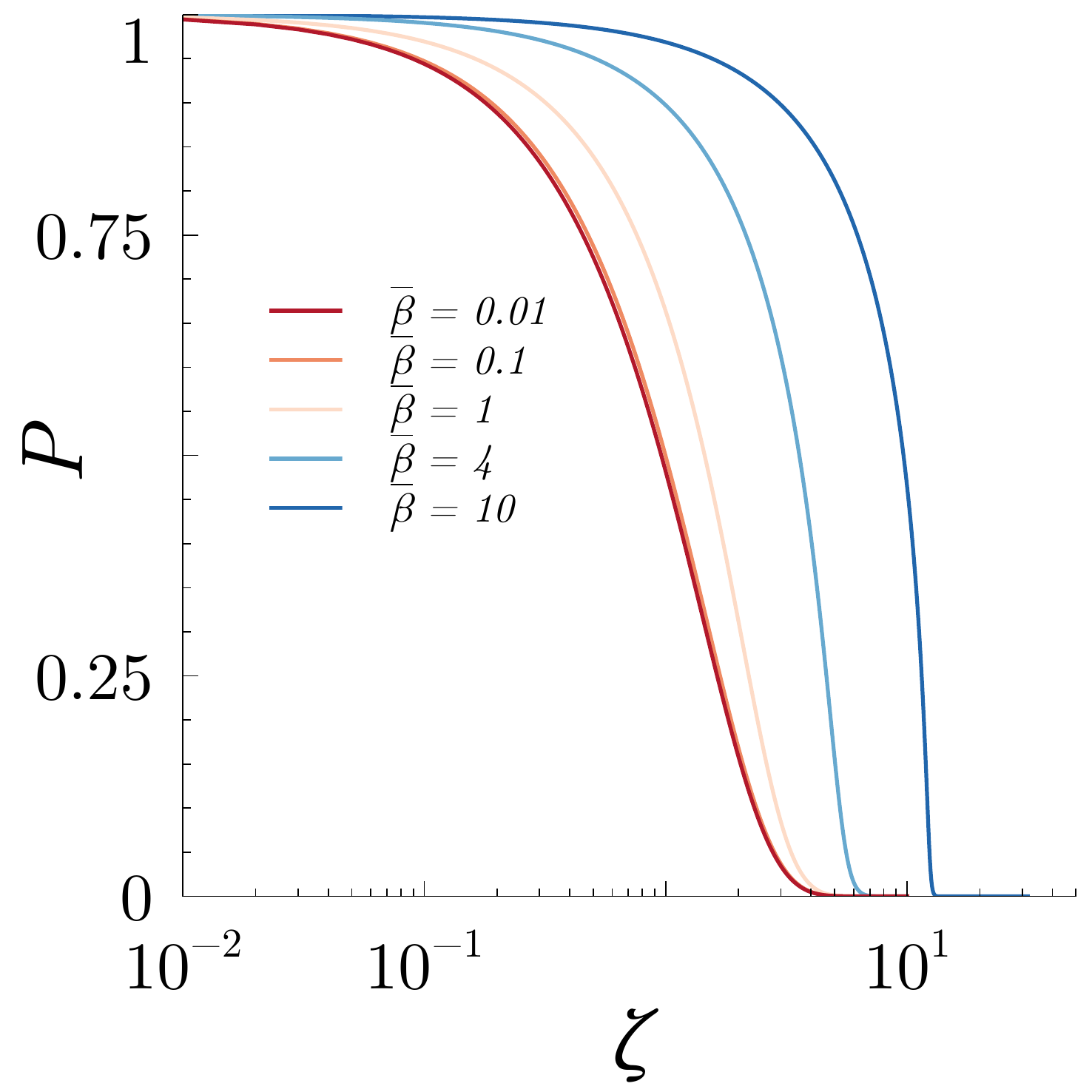} &
\includegraphics[width=0.5\textwidth]{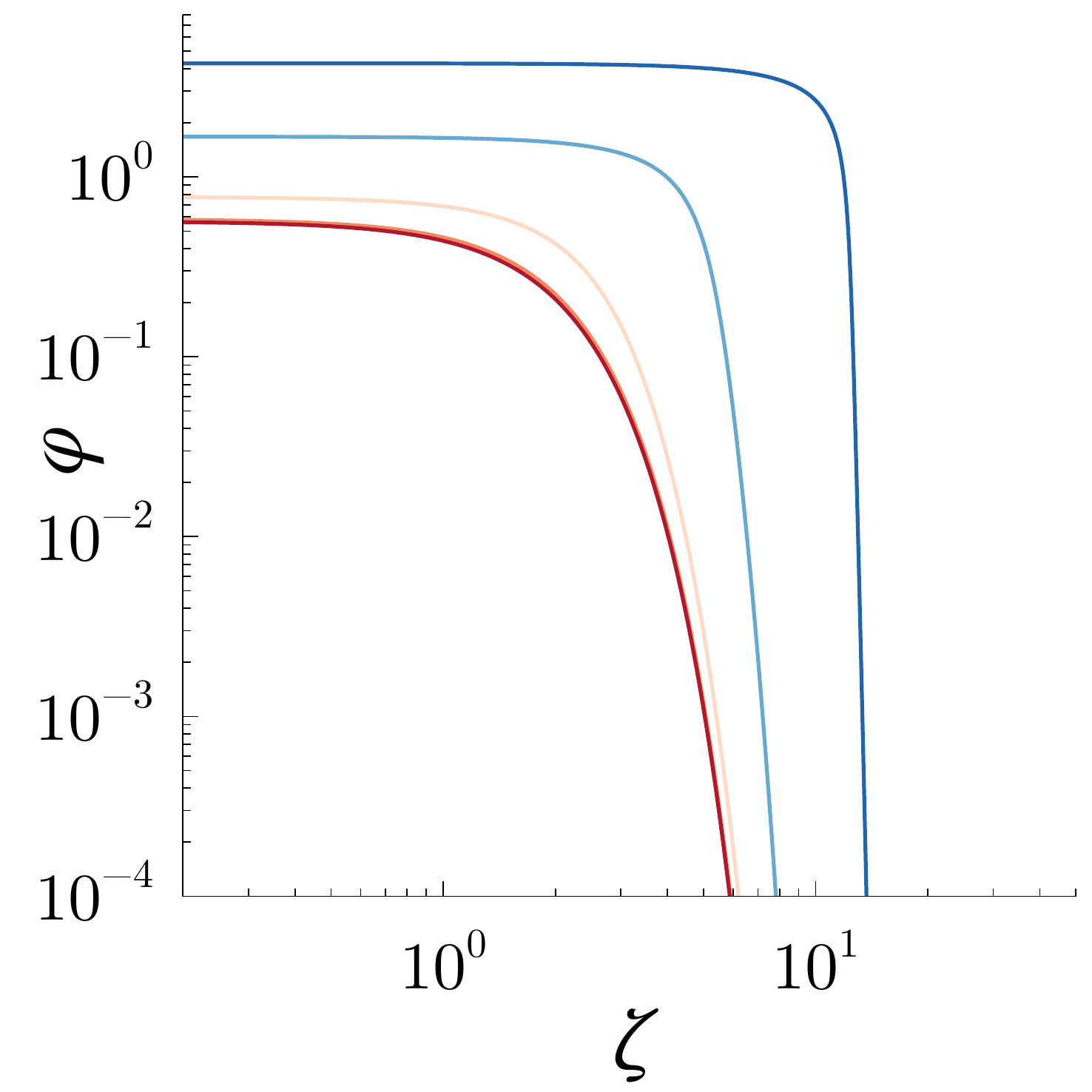}
\end{tabular}
\caption{(Colour online) Self-similar solution for the pressure-controlled case, showing ($a$) the pressure $P(\zeta)$ and ($b$) the rescaled flow rate $\varphi(\zeta)=Q\sqrt{T}$, for different values of $\tilde{\beta}$ indicated in the legend.\label{fig:fig5}}
\end{figure}

Under pressure-controlled conditions, equation~\eqref{eq:P_eq} admits an exact self-similar solution $P=P(\zeta)$ in terms of a rescaled coordinate $\zeta = z/\sqrt{T}$ at early times $T \ll 1$ before the diffusion front has reached the end of the channel. The profile $P(\zeta)$ satisfies the ordinary differential equation and boundary conditions 
\begin{equation}\label{eq:self_similar}
2(F(\tilde{\beta}P)\,P')'+ \zeta P'=0, \quad P(\zeta = 0) = 1, \quad \text{and} \quad P(\zeta \to \infty) \to 0,
\end{equation}
where primes indicate derivatives with respect to $\zeta$. Note from~\eqref{eq:flow_rate} that the corresponding flow rate is of the form $Q = \varphi(\zeta)/\sqrt{T}$, and the result $Q_0 = \varphi(0)/\sqrt{T}$ is in agreement with the early-time behaviour of the numerical results shown in figure~\ref{fig:fig3}($a$) below. Figure~\ref{fig:fig5} shows $P$ and $\varphi$ as a function of $\zeta$ for several values of $\tilde{\beta}$ indicated in the legend. As $\tilde{\beta}$ increases (e.g.\ bending modulus decreases), the displacement of the top wall increases and hence a higher pressure and flow rate are achieved within the channel.

For $\tilde{\beta} \ll 1$, the self-similar equation~\eqref{eq:self_similar} becomes the classical diffusion equation with a constant input $P=1$, with solution
\begin{equation}\label{eq:selfsimilar_beta_small}
P(\zeta) = \text{erfc} \left( \frac{\zeta}{2} \right), \qquad Q(\zeta) = \frac{e^{-(\zeta/2)^2}}{\sqrt{\upi T}}.
\end{equation}
Equation~\eqref{eq:selfsimilar_beta_small} fits the curves of figure~\ref{fig:fig5} when $\tilde{\beta} \ll 1$ and also the early-time trend $Q_0 = 1/\sqrt{\upi T}$ observed in figure~\ref{fig:fig3}($a$). It also agrees with the early-time behaviour of the small-$\tilde{\beta}$ asymptotic solution~\eqref{eq:pc_P0} derived in appendix~\ref{app:asymp}.

Moreover, for $\tilde{\beta} \gg 1$, using the approximation $F(\tilde\beta P) \approx (250/3003)\tilde \beta^3 P^3$, the equations~\eqref{eq:P_eq} and~\eqref{eq:self_similar} become (after a rescaling) the well-known equations for a viscous gravity current and their self-similar form, which have been solved by~\citet{Huppert1982}.

Although a flow-rate-controlled configuration does not admit an early-time self-similar solution for general $\tilde{\beta}$, it does so in the limits $\tilde{\beta} \to 0$ and $\tilde{\beta} \to \infty$. First, when $\tilde{\beta} \to 0$, equation~\eqref{eq:P_eq} again becomes the diffusion equation but with constant input flux $-\partial_Z P = 1$, whose self-similar solution is given by
\begin{equation}
P(\zeta) = \sqrt{T} \int_{\zeta}^{\infty} \text{erfc} \left( \frac{s}{2} \right) \text{d} s, \quad Q(\zeta) = \text{erfc}\left(\frac{\zeta}{2}\right).
\end{equation}
The requirement that $F(\tilde\beta P) \approx 1$ yields the condition $T \ll \tilde \beta^{-2}$ for validity. This result also agrees with the small-$\tilde\beta$ asymptotic solution~\eqref{eq:fr_P0} derived in appendix~\ref{app:asymp}, and the result $P_0 = 2\sqrt{T/\upi}$ agrees with the early-time behaviour of figure~\ref{fig:fig3}($b$) below. In the opposite limit, $\tilde{\beta} \to \infty$, we again obtain the gravity-current equation but with constant input flux, for which the self-similar solution has $Z \sim \tilde{\beta}^{3/5} T^{4/5}$ and $P \sim \tilde{\beta}^{-3/5} T^{1/5}$~\citep{Huppert1982}, and the conditions for validity $\tilde\beta P \gg 1$ and $Z \ll 1$ yield $\tilde\beta^{-2} \ll T \ll \tilde{\beta}^{-3/4}$.

\section{Results}
\label{sec:results}

This section is devoted to presenting the results obtained with the quasi-steady lubrication theory developed in~\S\ref{subsec:quasi_steady}, and to compare these results with those extracted from the 3D numerical simulations described in~\S\ref{sec:3D_numerical_simulations}. The quasi-steady lubrication equation~\eqref{eq:P_eq} was solved numerically with a standard finite-difference scheme. In particular, we have computed the fluid pressure distribution $P(Z,T)$, the associated flow-rate distribution, $Q(Z,T)$, and the vertical wall displacement, $U_Y(X,Z,T)$, for several values of $\tilde{\beta}$. As shown by~\cite{Christov2018}, higher flow rates are achieved as $\tilde{\beta}$ increases, corresponding to a larger upper wall deformation and a reduced hydraulic resistance. For instance, in a steady pressure-controlled situation, the dimensionless flow rate is higher for a given pressure drop~\citep{Christov2018}, 
\begin{equation}
Q \left(T \to \infty\right) \to 1 + \frac{1}{2} \tilde{\beta} + \frac{10}{63} \tilde{\beta}^2 + \frac{125}{6006} \tilde{\beta}^3,
\end{equation}
since the cross-section increases as the top wall deforms. This result can be observed in figure~\ref{fig:fig3}($a$,$b$), which shows the dependence of the inlet flow rate in a pressure-controlled configuration, $Q_0=Q(Z=0,T)$ in ($a$), and the inlet overpressure $P_0 = P(Z=0,T)$ in flow-rate-controlled conditions in ($b$), as a function of time $T$ for several values of $\tilde{\beta}$. The insets show how the solution of~\cite{Christov2018} (dashed lines) is reached for $T\gtrsim O(1)$, revealing that $t_c$ is indeed the proper time scale. In pressure-controlled configurations $Q_0$ decreases with time as the initial condition for the fluid pressure diffuses along the channel and $\partial_Z P (Z=0,T)$ decreases in magnitude. In flow-rate-controlled conditions, $P_0$ increases with time since the magnitude of $\partial_Z P$ decreases as the fluid spreads, so that $F(\tilde{\beta}P)$ must increase to keep $Q_0 =1$.

To illustrate the diffusion of $P(Z,T)$ and $Q(Z,T)$, and the displacement $\beta U_Y(X=0,Z,T)$ along the channel, we show in figure~\ref{fig:fig6} (dashed lines) their time evolution for $\tilde{\beta} = 0.4$, under pressure-controlled conditions ($a$) and flow-rate-controlled conditions ($b$). Here, $U_Y(X=0,Z,T) = U_m(Z,T) = P(Z,T)/384$ is the maximum displacement of the top wall according to equation~\eqref{eq:displacement}. The results obtained at the largest time indicated in the legend correspond to the steady solution of~\cite{Christov2018}.

\begin{table}
\centering
\setlength{\tabcolsep}{6pt}
\begin{tabular}{c c c c c c c c c c c c c c c c}
 & $h_0$ & $w$ & $\ell$  & $d$  & $B$  & $\nu$ & $\rho_s$ & $\Delta p$ & $\tilde{\beta}$\\
 & (\text{mm}) &  (\text{mm}) & (\text{mm}) & (\text{mm}) & (\text{$\mu$J}) & & (\text{kg m}$^{-3}$) & (\text{kPa}) & \\
\textbf{S4} & 0.244 &  1.7 & 15.5 & 0.2 & $\approx$ 1.6 & 0.499 & 970 & 4.487 & 0.4 \\
\hline
\end{tabular}
\begin{tabular}{lcccccc}
& $\rho$ & $\mu$   & $q_0$ & $\dfrac{\epsilon Re}{12}$  & $\dfrac{\epsilon Re}{4 \tilde{\beta}}$ & $\tilde{\gamma}$\\
& (\text{kg m}$^{-3}$) & (\text{Pa s})  & (\text{ml min}$^{-1}$) &   &  \\
\textbf{Case I}  & 970 &  0.485  & 7.38 $\times$ 10$^{-2}$  & 1.9 $\times$ 10$^{-6}$ & 1.42 $\times$ 10$^{-5}$ & 2.1 $\times$ 10$^{-6}$  \\
\textbf{Case II}  & $10^{3}$ & $10^{-3}$  & 35.76 & 0.46 & 3.45 & 0.49 
\end{tabular}
\caption{\label{tab:table1} Physical parameters corresponding to the S4 experiment of~\cite{Ozsun2013} for two different working liquids, namely water (II), which is the least favorable configuration, and a silicon oil of 500 cSt (I), for which the assumptions of~\S\ref{subsec:quasi_steady} are satisfied. The values of $q_0$ and $\Delta p$ ensure that $\tilde{\beta} = 0.4$, both in the pressure-controlled case and in the flow-rate-controlled situation.}
\end{table}

\subsection{Start-up time and comparison with previous models}\label{subsec:start_up_time}
\begin{figure}
\centering
\begin{tabular}{cc}
($a$) & ($b$)  \\
\includegraphics[width=0.45\textwidth]{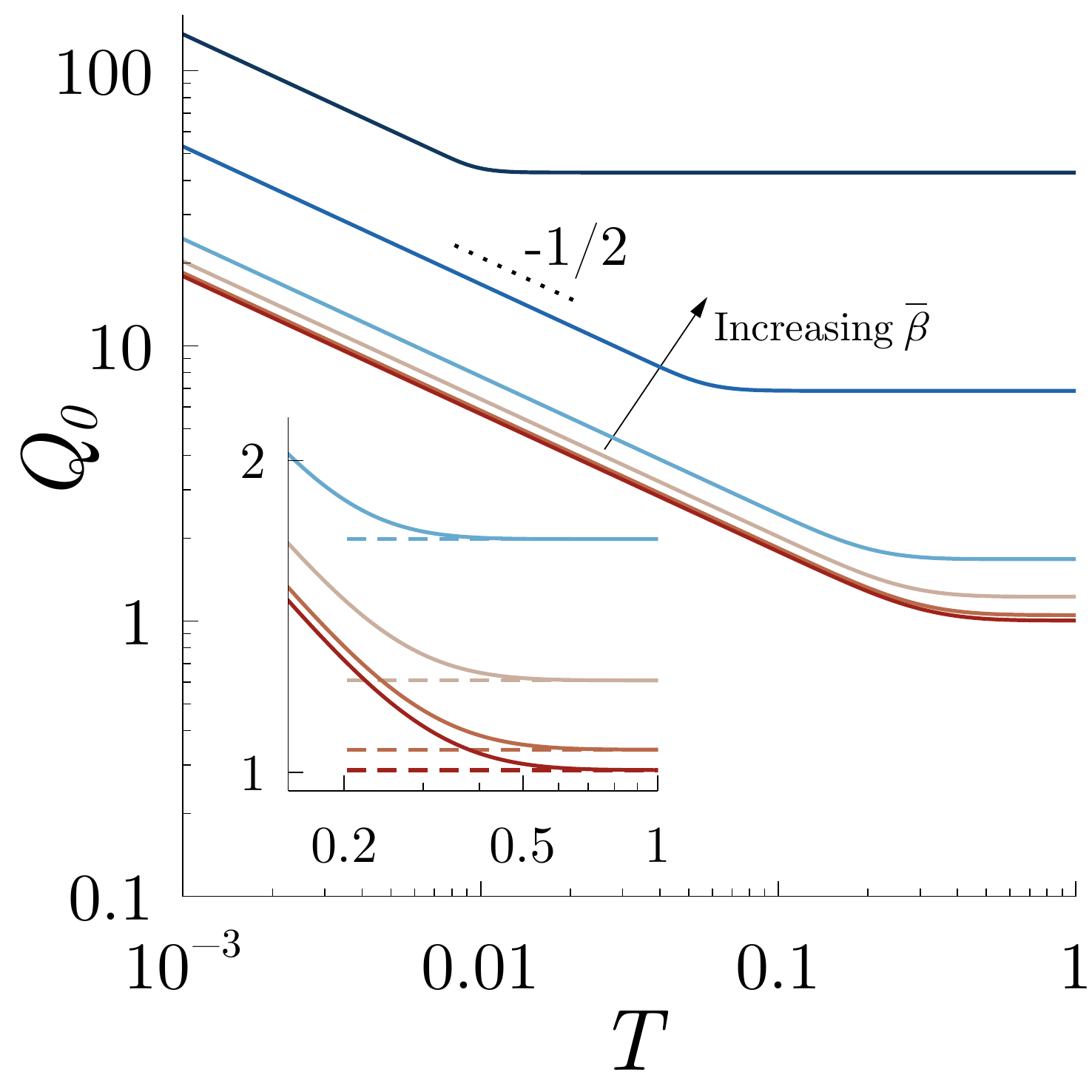}  & \includegraphics[width=0.45\textwidth]{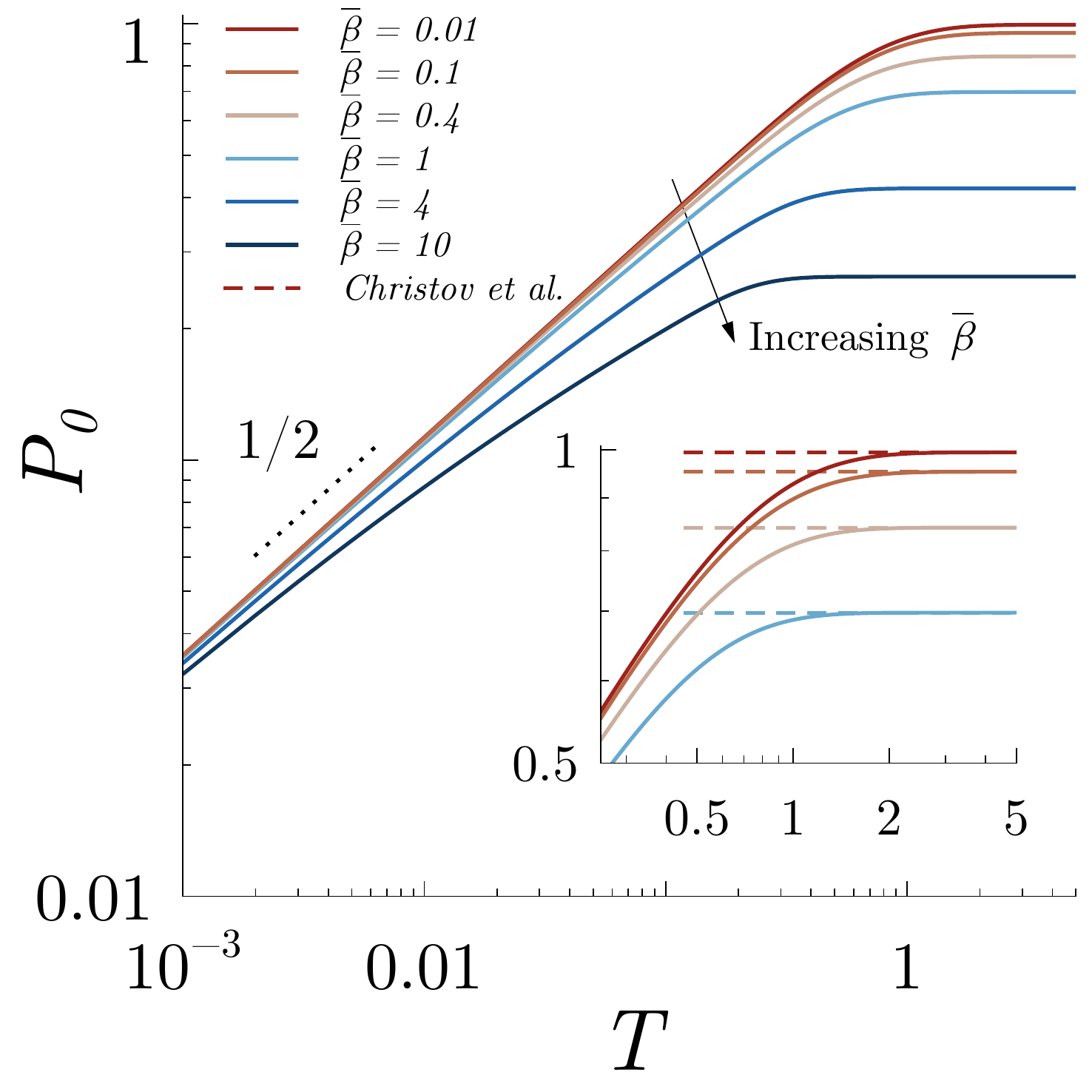}\\
($c$) & ($d$)\\
\includegraphics[width=0.45\textwidth]{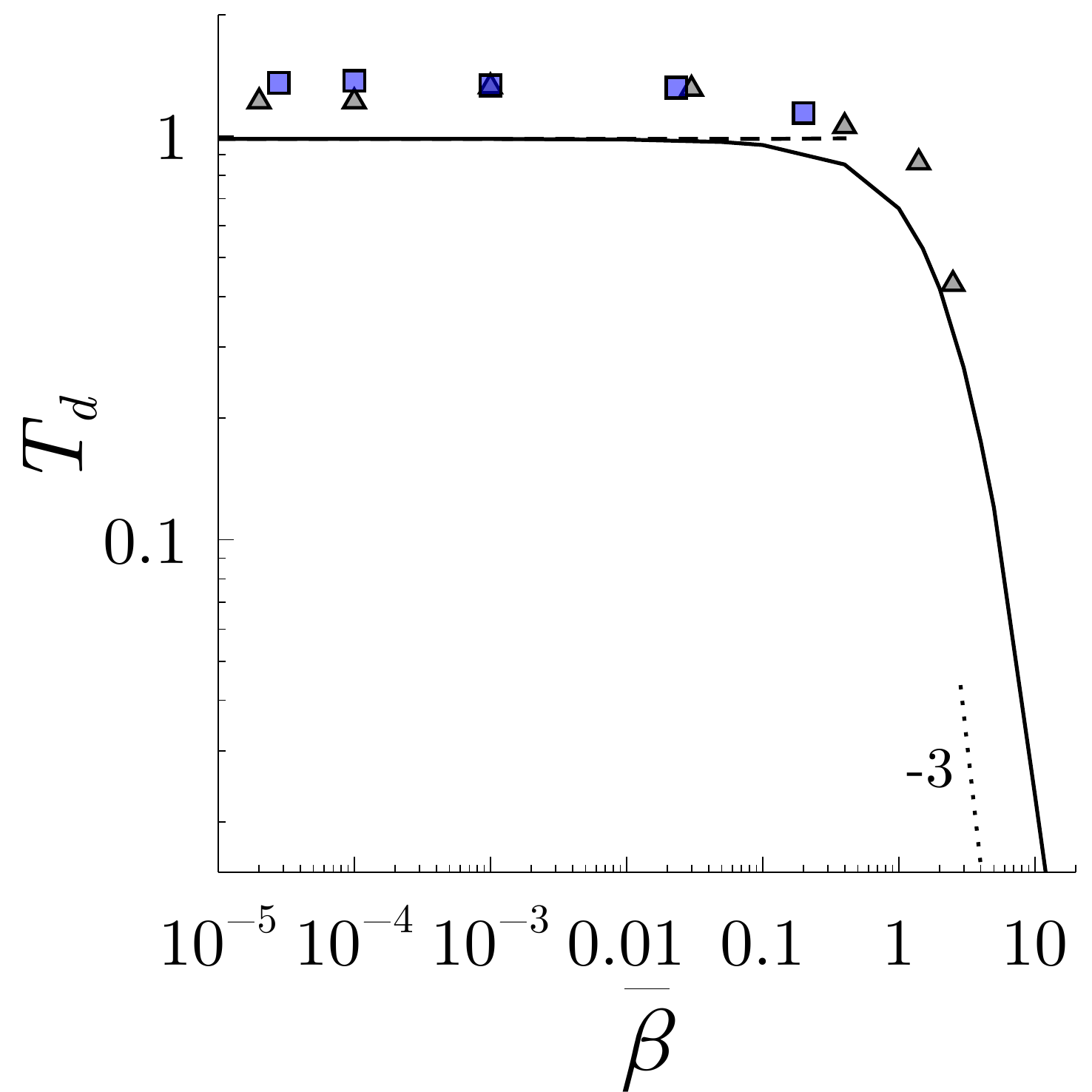}  & \includegraphics[width=0.45\textwidth]{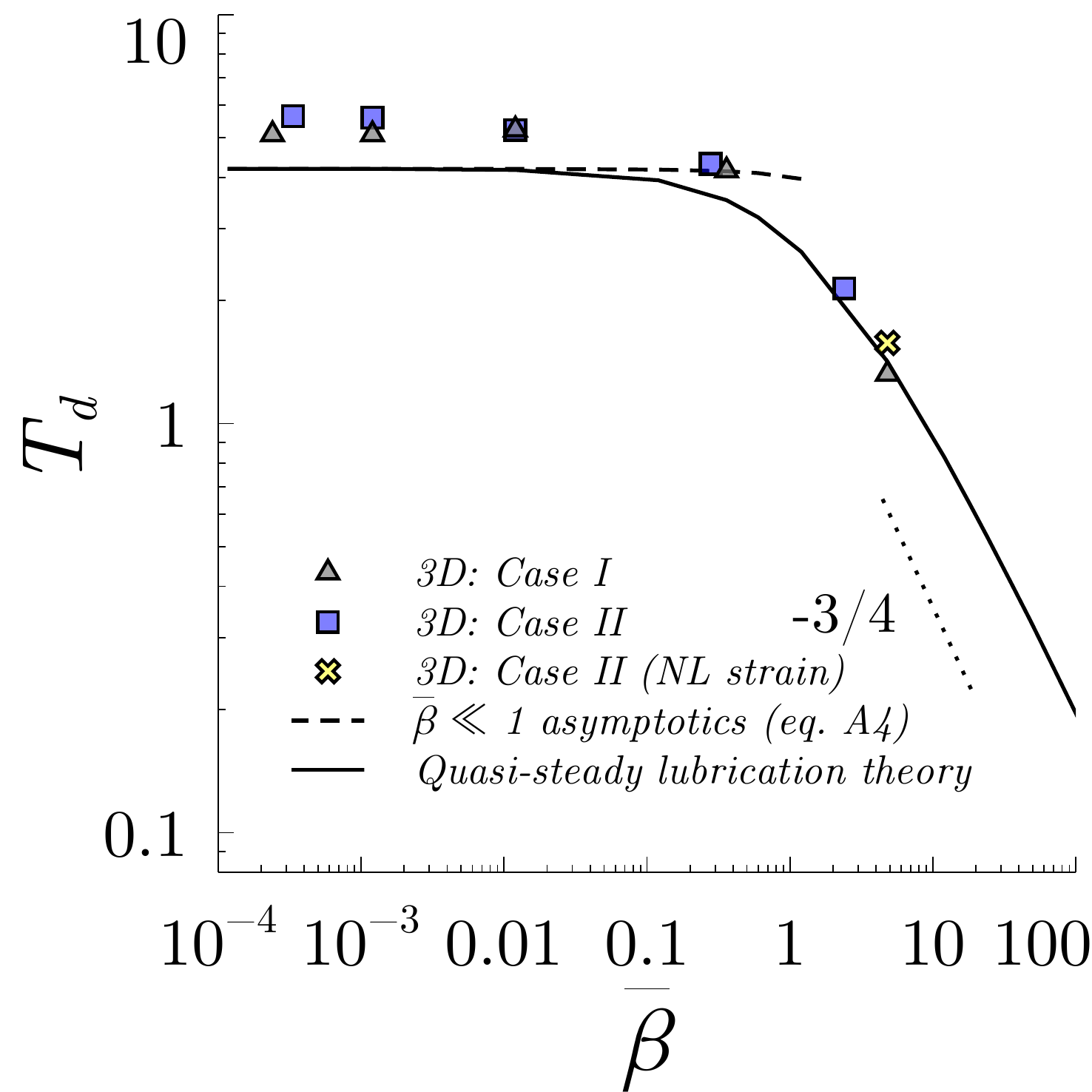}
\end{tabular}
\caption{(Colour online) ($a$) Inlet flow rate $Q_0$ for a pressure-controlled channel and ($b$) inlet overpressure $P_0$ in a flow-rate-controlled configuration, as a function of time $T$, for several values of $\tilde{\beta}$ as indicated in the legend. The insets show the solution of~\cite{Christov2018} reached at long time (dashed lines). The start-up time $T_d$ is shown for ($c$) a pressure-controlled configuration and for ($d$) a flow-rate-controlled configuration. The solid lines correspond to the quasi-steady lubrication theory given by equations~\eqref{eq:P_eq}--\eqref{eq:c}. The symbols correspond to the 3D numerical simulations for two combinations of the dimensionless parameters $\epsilon Re$ and $\tilde{\gamma}$ reported in table~\ref{tab:table1}, and the dashed lines correspond to the asymptotic solution for $\tilde{\beta} \ll 1$ obtained in appendix~\ref{app:asymp}.\label{fig:fig3}}
\end{figure}

\begin{figure}
\centering
\begin{tabular}{cc}
($a$) & ($b$)\\
 \includegraphics[width=0.5\textwidth]{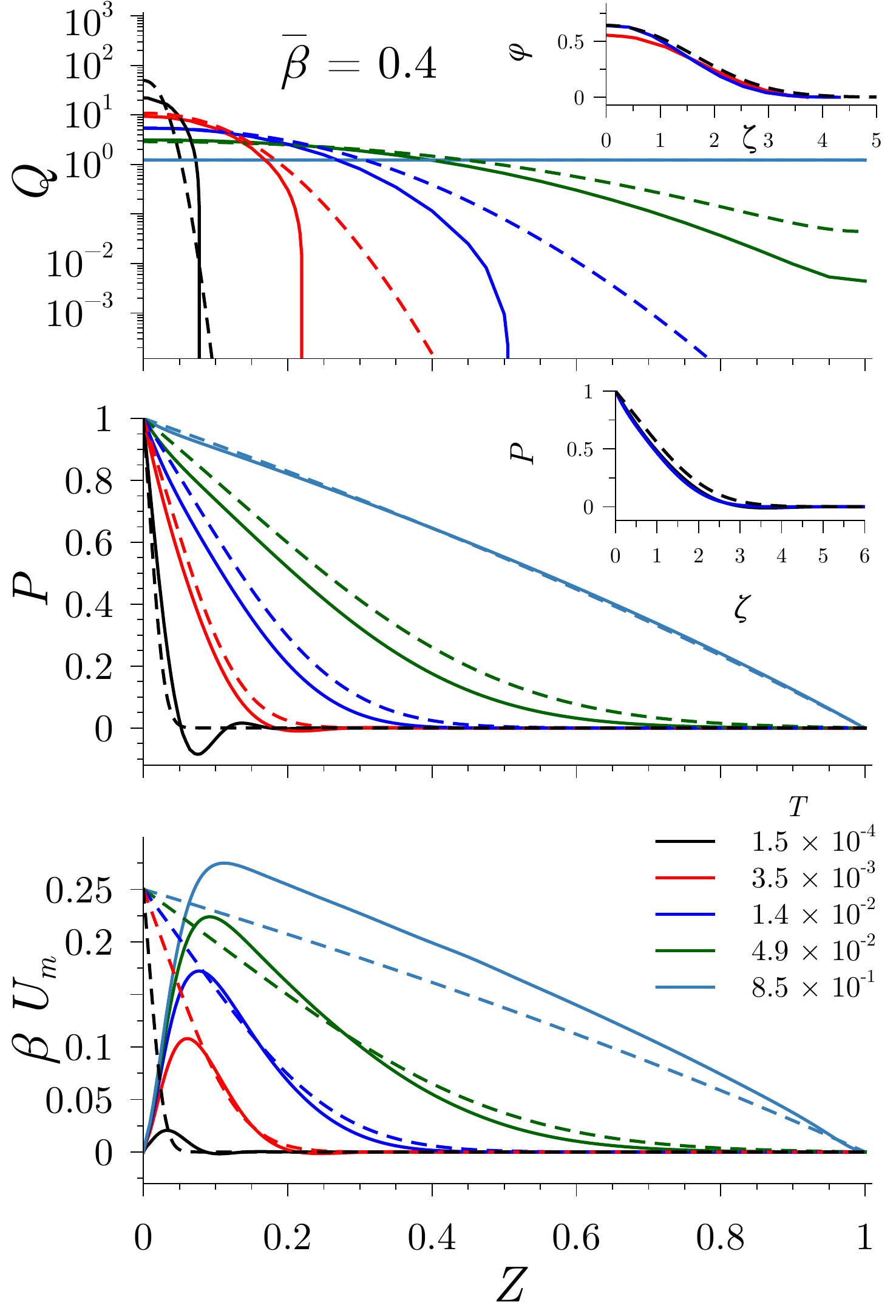} &
\includegraphics[width=0.5\textwidth]{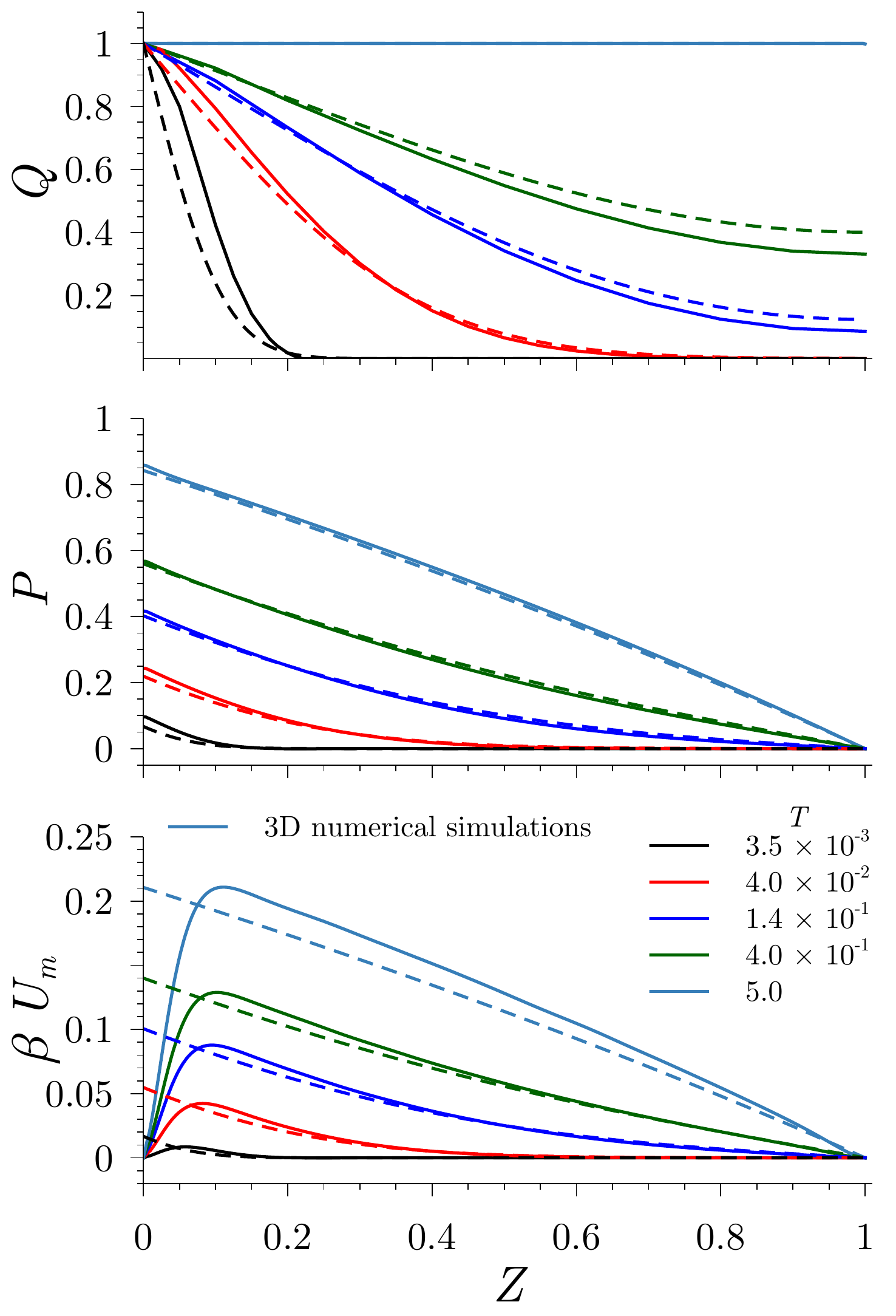}
\end{tabular}
\caption{(Colour online) Flow rate $Q(Z,T)$, pressure field $P(Z,T)$, and maximum displacement $\beta U_m$, as functions of $Z$ for $\tilde{\beta} = 0.4$ at different times indicated in the legend. The system is pressure-driven in ($a$), and flow-rate controlled in ($b$). The dashed lines show the lubrication solution, and the solid lines the 3D numerical simulations. Here, $\epsilon\,Re/12 \simeq 1.9\times 10^{-6}$, $\epsilon Re/(4 \tilde{\beta}) \simeq 1.42\times 10^{-5}$ and $\tilde{\gamma}\simeq 2.1\times 10^{-6}$, fulfilling the lubrication hypotheses. The insets show the self-similar solution given by equation~\eqref{eq:self_similar}, together with the rescaled numerical solution for two different times near start-up.\label{fig:fig6}}
\end{figure}

The start-up time at which the steady-state solution of~\cite{Christov2018} is reached, denoted by $T_d$, is defined here by the condition $|\partial_T Q(Z = 1,T_d)| = 10^{-4}$. We have ensured that the value of $T_d$ is robust against alternative choices of the dependent variable. For instance, $\max_Z |\partial_T P(Z,T_d) | = 10^{-4}$ provides very similar values. The start-up time is shown in figures~\ref{fig:fig3}($c$,$d$) as a function of $\tilde{\beta}$ in a pressure-controlled and in a flow-rate controlled configuration, respectively, computed with the quasi-steady lubrication model~\eqref{eq:P_eq}, with the 3D numerical simulations of~\S\ref{sec:3D_numerical_simulations}, and with the small-deformation ($\tilde\beta \ll 1$) asymptotic solutions~\eqref{eq:asymp_P0} from appendix~\ref{app:asymp}. Figures~\ref{fig:fig3}($c$,$d$) reveal that the start-up time decreases as the compliance increases. For instance, the value of $T_d$ decreases by a factor of 2 when $\tilde{\beta}$ is increased from $0$ to $1.6$ (pressure-controlled) or $2$ (flow-rate-controlled). In the limit $\tilde{\beta}\to 0$, the values of $T_d$ reach the corresponding rigid-channel asymptotes, $T_d\simeq 0.98$ and $T_d\simeq 4.2$, when $\tilde{\beta} \lesssim 0.02$ and $\tilde{\beta} \lesssim 0.01$, respectively. Consequently, the asymptotic solutions developed in~\S\ref{app:asymp} for $\tilde{\beta}\ll 1$ are valid for $\tilde{\beta} \lesssim 0.01$. In the opposite limit, $\tilde{\beta} \gg 1$, we can obtain the scalings $T_d \propto \tilde{\beta}^{-3}$ for a pressure-controlled configuration, and $T_d \propto \tilde{\beta}^{-3/4}$ for a flow-rate-controlled situation, from the gravity-current solutions deduced in~\S\ref{subsec:self_similar}, or by setting $u_c \sim h_0$ in the scaling arguments of~\S\ref{subsec:time_scales}.

The results of the 3D numerical simulations shown in figure~\ref{fig:fig3}($c$,$d$) were computed for the geometry of the S4 experiment of~\cite{Ozsun2013}, different values of $\tilde{\beta}$, and two different combinations of $\epsilon Re$ and $\tilde{\gamma}$, corresponding to two different working liquids (see table~\ref{tab:table1}). Case I corresponds to a silicon oil of viscosity 500 cSt as working liquid, where $\epsilon Re/12 = 1.9 \times 10^{-6}$ and $\tilde{\gamma} = 2.1 \times 10^{-6}$ (triangles), and Case II corresponds to water, which is the least favourable case for this S4 geometry since $\epsilon Re/12 = 0.46$ and $\tilde{\gamma} = 0.49$ (squares), and thus the inertia of the liquid and of the solid may have influence on the flow. However, we have found fair agreement between the quasi-steady lubrication model with negligible solid and liquid inertia and the 3D numerical simulations in both cases. In particular, the agreement improves for increasing values of $\tilde{\beta}$, indicating that the relative importance of the solid and liquid inertia becomes smaller for larger wall displacements. In particular, the local liquid inertia becomes negligible as $\tilde{\beta} \to\infty$, since $\epsilon Re/(4 \tilde{\beta}) \to 0$. Morever, $\tilde{\gamma}$ can be expressed in terms of $\tilde{\gamma} \propto \rho_s d h_0^2 p_c \epsilon^3/(12 \mu^2 \ell \tilde{\beta})$, so that the solid inertia becomes negligible as $\tilde{\beta}$ increases for a fixed geometry and liquid.


We have also considered the lubrication model derived by~\cite{Dendukuri2007}, and later used by~\cite{Panda2009}, in which the spanwise average of the upper wall's displacement is assumed to be linearly proportional to the fluid pressure, and thereby the pressure and the displacement fields are only functions of time and the longitudinal coordinate. This type of approximation is usually known as the Winkler foundation~\citep{Kerr1964}. To obtain the dimensionless version of their model we take the characteristic displacement as $u_c^{D} = w p_c/E$, and the characteristic time as $t_c^{D} = \mu \ell^2 w/(12 E h_0^3)$. Hence, the new dimensionless variables are $t = \mu \ell^2 w/(12 E h_0^3) \mathcal{T}$ and $u_y = (w p_c/E) \, \mathcal{U}_Y$, and the nonlinear diffusion equation for $P(Z,\mathcal{T})$ reads
\begin{equation}\label{eq:dendukuri}
\partial_{\mathcal{T}} P = \partial_Z[\partial_Z P(1+\mathscr{D}P)^3],
\end{equation}
subjected to the same boundary and initial conditions as~\eqref{eq:P_eq}, and where $\mathscr{D} = w p_c/(h_0 E)$ is the associated compliance parameter. Note that, when $\tilde{\beta} \ll 1$ and $\mathscr{D} \ll 1$,~\eqref{eq:P_eq} and~\eqref{eq:dendukuri} coincide at leading order: $\partial_T P = \partial^2_Z P + O(\tilde{\beta})$, or $\partial_{\mathcal{T}} P = \partial^2_Z P + O(\mathscr{D})$. However, since the scalings for the pressure and the displacement field are different from our lubrication model, the ratios between the different characteristic scales and the two compliance parameters depend on $\nu$ and on the ratio $w/d$, namely $t_c/t_c^D = 12/5 (1-\nu^2) (w/d)^3$, $u_c/u_c^D = 12(1-\nu^2)(w/d)^3$, and $\tilde{\beta}/\mathscr{D} = (1-\nu^2)/20 (d/w)^3$. Taking the S4 experiment of~\cite{Ozsun2013} with water as the working liquid (see table~\ref{tab:table1}), and considering $\tilde{\beta} = 0.4$ as a typical configuration, then $\mathscr{D} = 0.017$, and the dimensional steady-state times predicted by each model are $t_d = 1.22$ ms (present work) and $t_d = 1.42$~$\mu$s~\citep{Dendukuri2007,Panda2009}, whose ratio is of the same order as $t_c/t_c^D$ for the parameter values of Case I in table~\ref{tab:table1}. Hence, there is a strong quantitative disagreement between the unsteady lubrication model of~\cite{Dendukuri2007}, with both our lubrication theory, and the 3D numerical simulations. We thus conclude that the model of~\cite{Dendukuri2007} fails to predict the transient flow, especially in microchannels where the thickness of the top wall is smaller than, or of the same order as, the channel width. This situation resembles the shortcomings found in previous steady lubrication models~\citep{Gervais2006,Hardy2009,Cheung2012,Raj2016,Raj2017}, in that the fitting parameters that appear in these model have been used even for microchannels with thin upper walls, as pointed out by~\cite{Christov2018}. Indeed, just like these fitting parameters, $\mathscr{D}$ naively absorbs the geometric and material constants.

\subsection{Transient comparison between the quasi-steady lubrication theory and the 3D numerical simulations}\label{subsec:comparison}

To perform a more detailed comparison between the results of the quasi-steady lubrication model~\eqref{eq:P_eq}--\eqref{eq:c} and the 3D simulations, we have chosen the S4 experiment of~\cite{Ozsun2013}, whose fixed geometrical and physical parameters are reported in table~\ref{tab:table1}. We have also taken $\tilde{\beta} = 0.4$ as a typical value of the compliance parameter, and thus $\Delta p$ and $q_0$ (or $V_{\ell}$) are chosen to ensure this value in a pressure-controlled and in a flow-rate-controlled situation, respectively. In the experimental results reported by~\cite{Ozsun2013} water was used as working liquid, which corresponds to Case II of table~\ref{tab:table1} and is an unfavourable case that does not fulfill the lubrication assumptions of~\S\ref{subsec:quasi_steady}. Note that the latter case corresponds to one of the results of the steady 3D simulations shown in figure~\ref{fig:fig2}($a$), which is in excellent agreement with the experiments. We have also considered a more favourable case, namely Case I of table~\ref{tab:table1}, so that the hypotheses behind~\eqref{eq:P_eq}--\eqref{eq:c} are satisfied.

\begin{figure}
\centering
\begin{tabular}{cc}
($a$) & ($b$)\\
 \includegraphics[width=0.5\textwidth]{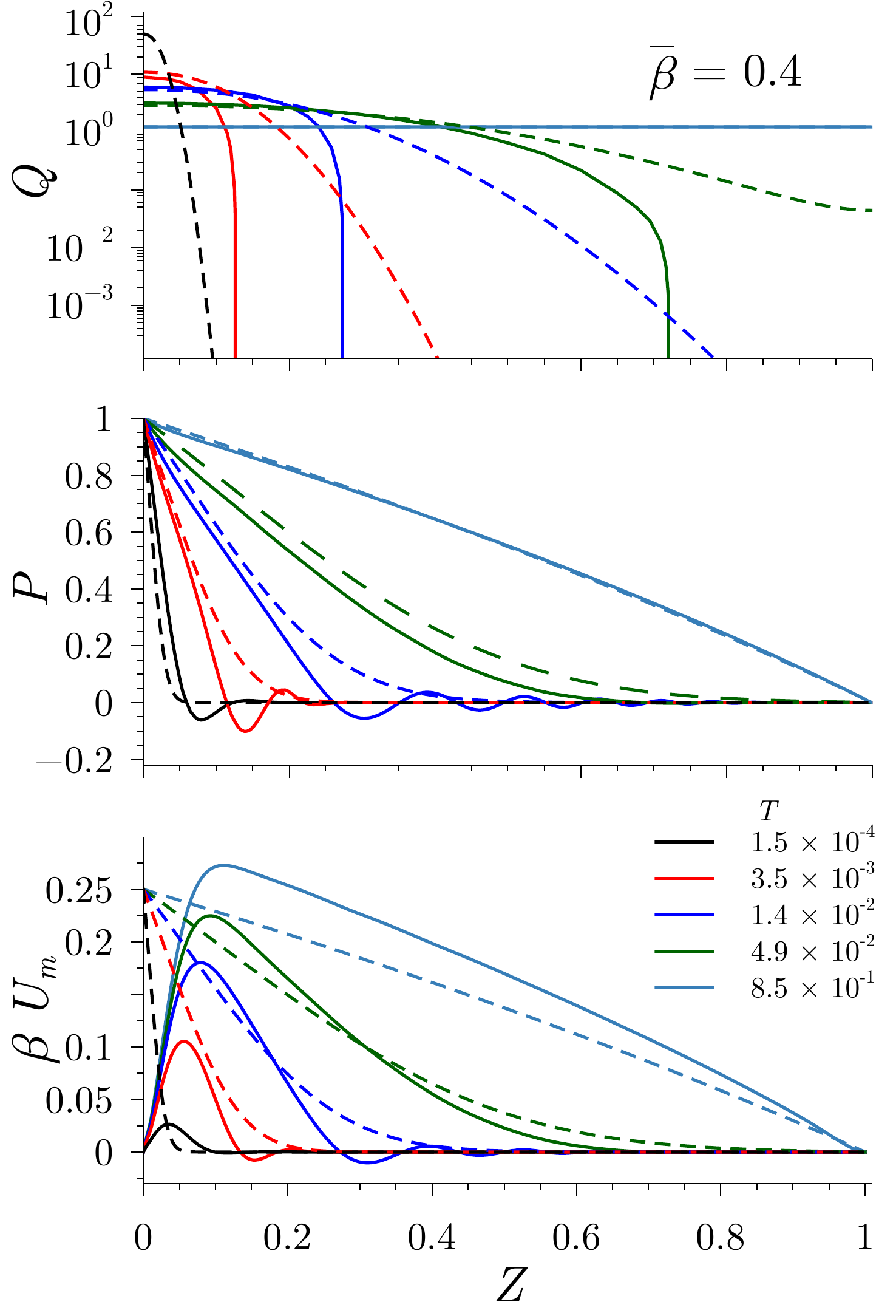} &
\includegraphics[width=0.5\textwidth]{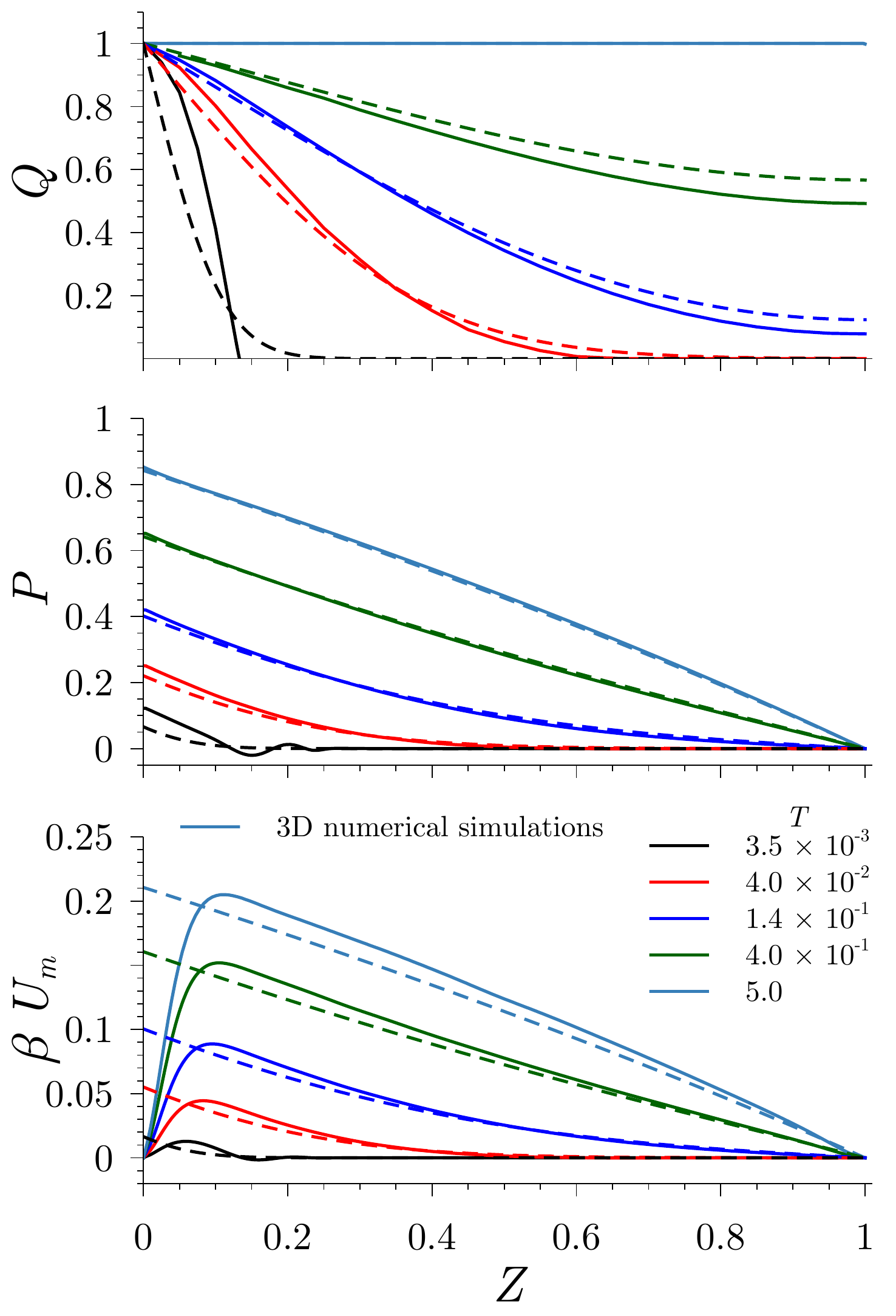}
\end{tabular}
\caption{(Colour online) Same as figure~\ref{fig:fig6}, but for $\epsilon\,Re/12 \simeq 0.46$, $\epsilon Re/(4\tilde{\beta}) \simeq 3.45$ and $\tilde{\gamma}\simeq 0.49$, corresponding to the conditions of the S4 experiment of~\cite{Ozsun2013}.\label{fig:fig7}}
\end{figure}

Figures~\ref{fig:fig6} and~\ref{fig:fig7} show the flow rate $Q$ (upper row), the pressure distribution $P$ (middle row) and the displacement of the wall $\beta U_m$ (bottom row) as functions of $Z$ at different times indicated in the legend, obtained from the lubrication theory (dashed lines) and from the 3D simulations (solid lines). In both figures, the configuration is pressure controlled in the left column, and flow-rate controlled in the right column. In the 3D simulations, the pressure drop is evaluated along the line $X = 0, Y = 1/2$, while the solid deformation field is evaluated at the fluid-solid interface and $X = 0$. The control parameters correspond to Case I in figure~\ref{fig:fig6} and to Case II in figure~\ref{fig:fig7}.

\begin{figure}
\hspace{-0.9cm}
\begin{tabular}{ccc}
($a$) & ($g$)  & ($m$) \\
  \includegraphics[width=0.36\textwidth]{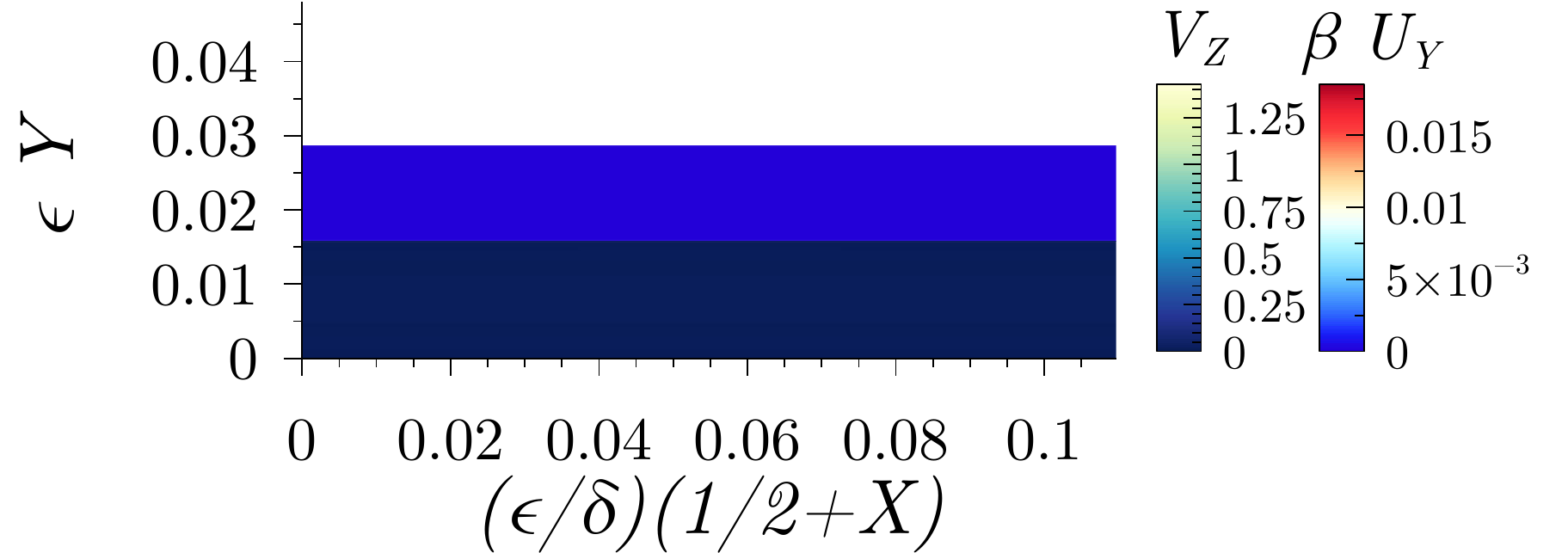} &   \includegraphics[width=0.36\textwidth]{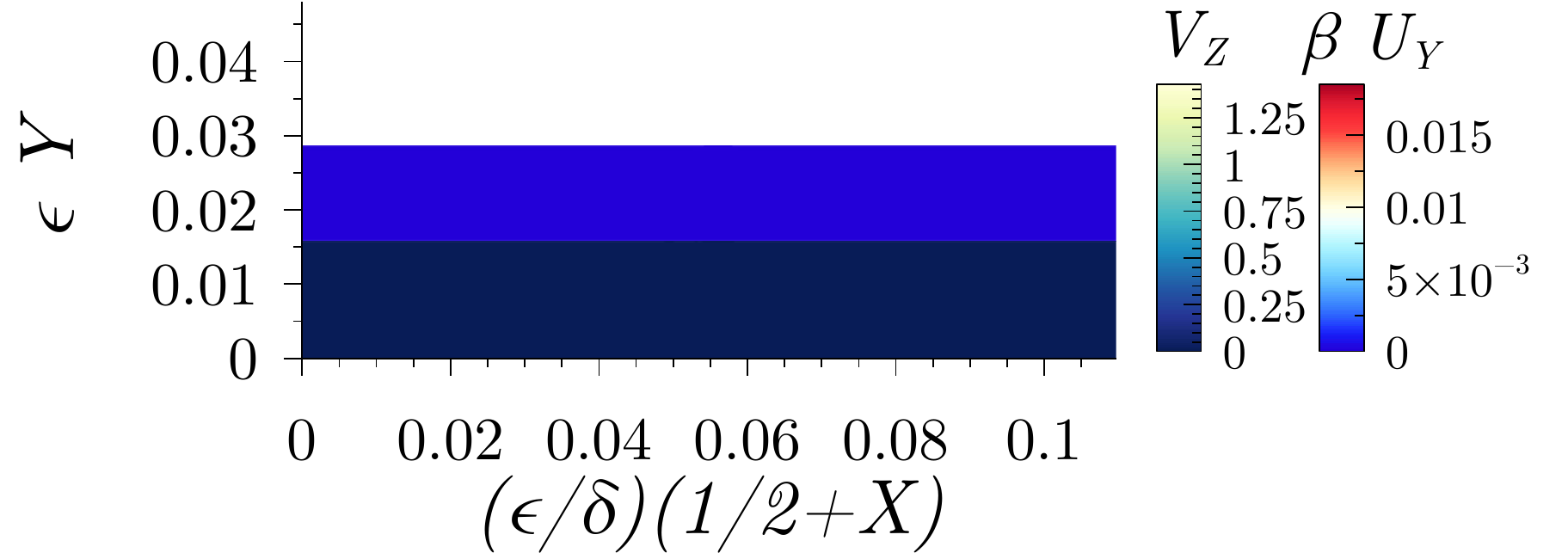} & \includegraphics[width=0.36\textwidth]{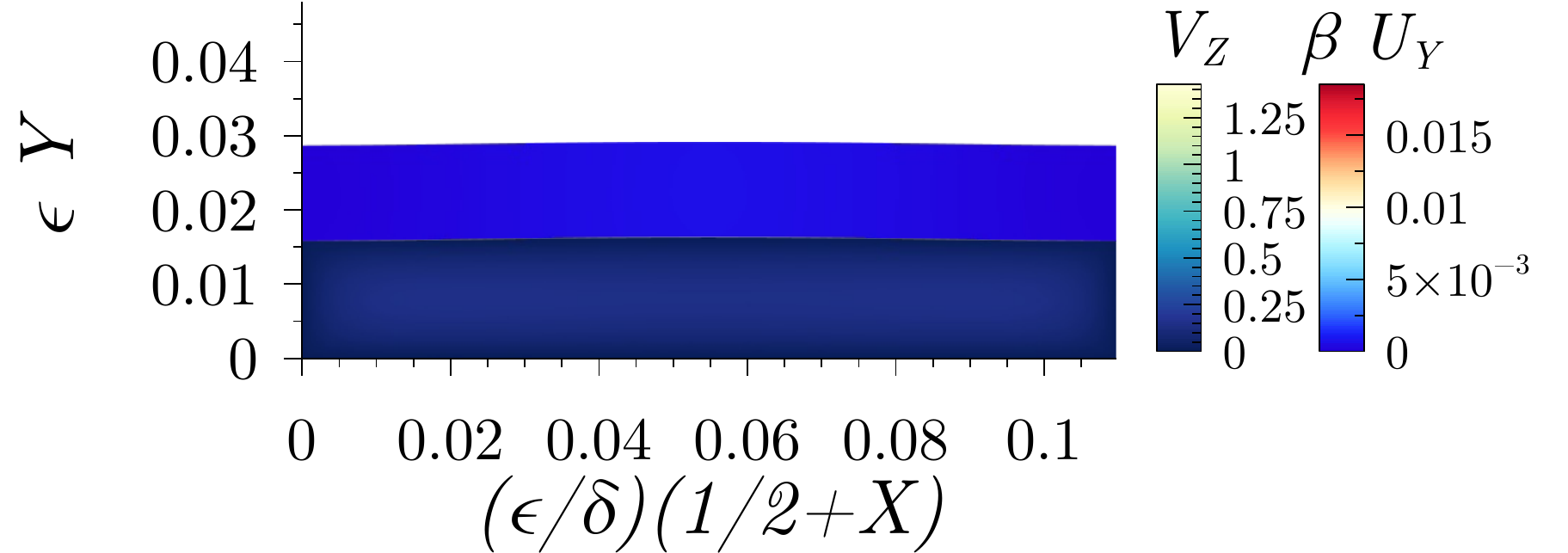}\\
  ($b$) & ($h$)  & ($n$)\\
\includegraphics[width=0.36\textwidth]{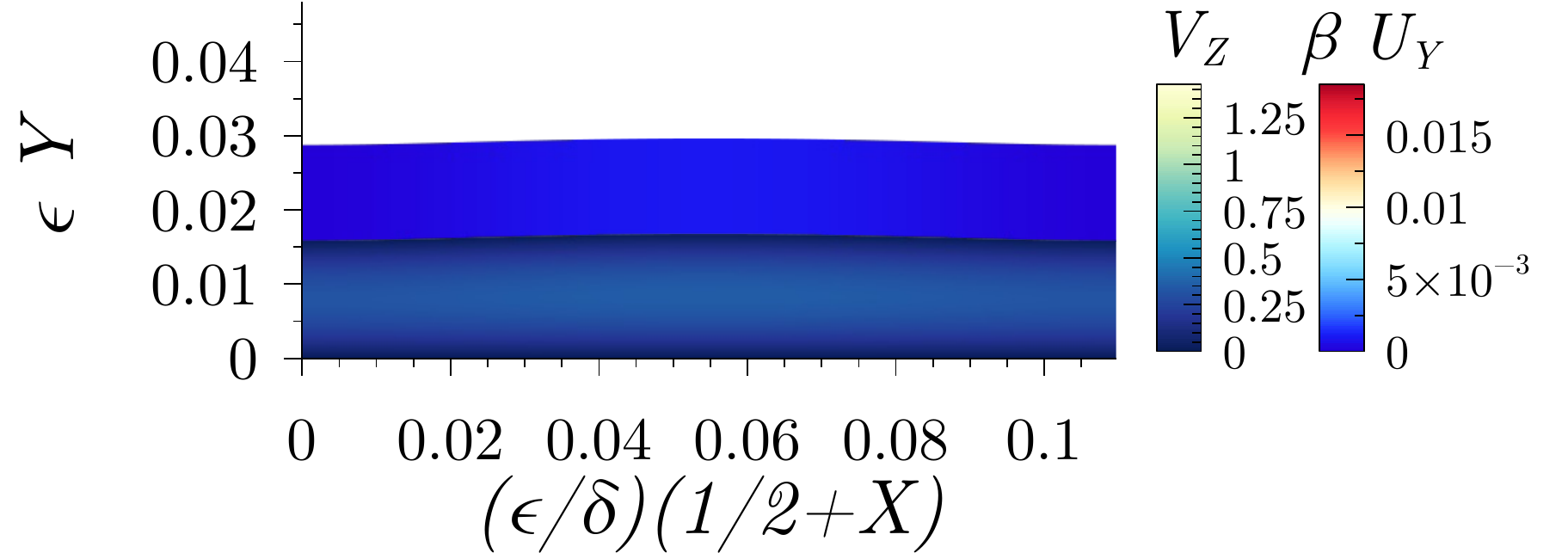} &   \includegraphics[width=0.36\textwidth]{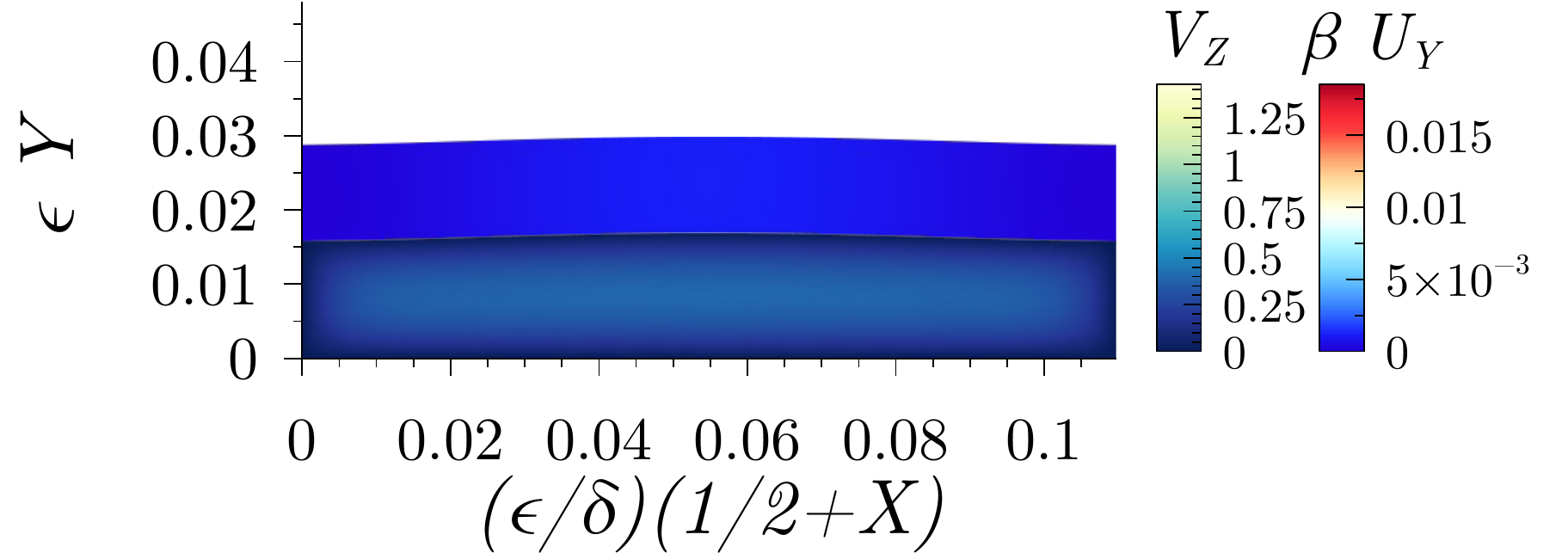} & \includegraphics[width=0.36\textwidth]{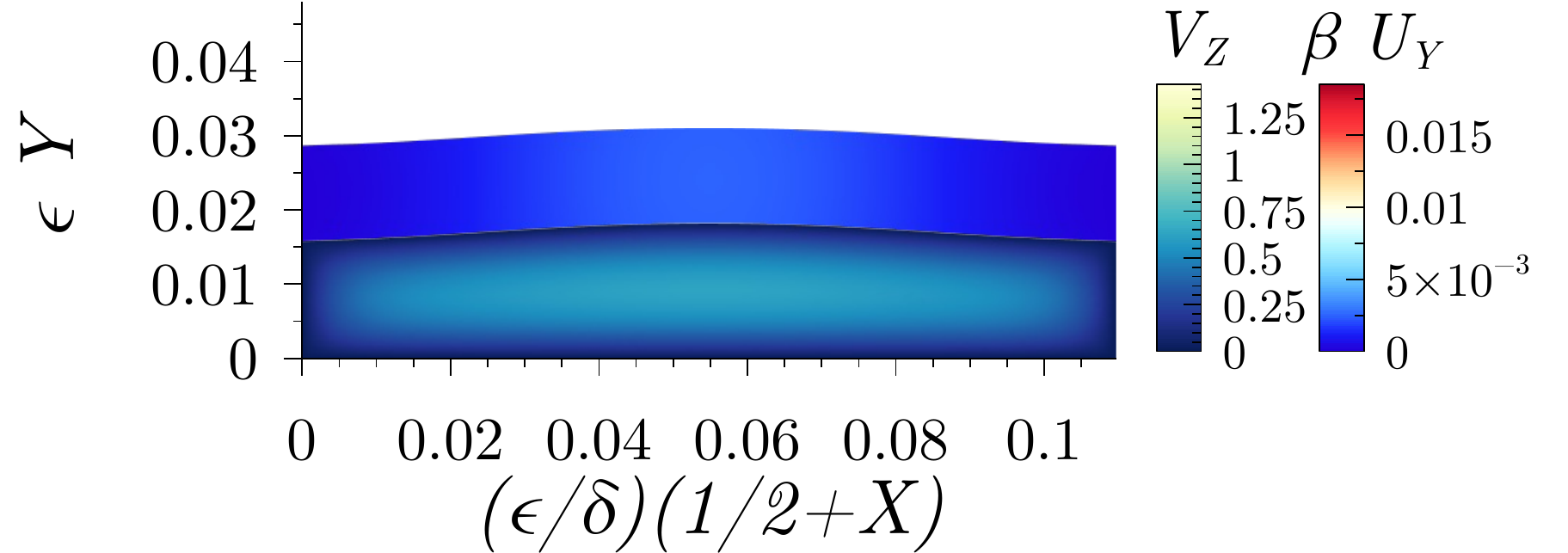} \\
($c$) & ($i$)  & ($o$) \\
   \includegraphics[width=0.36\textwidth]{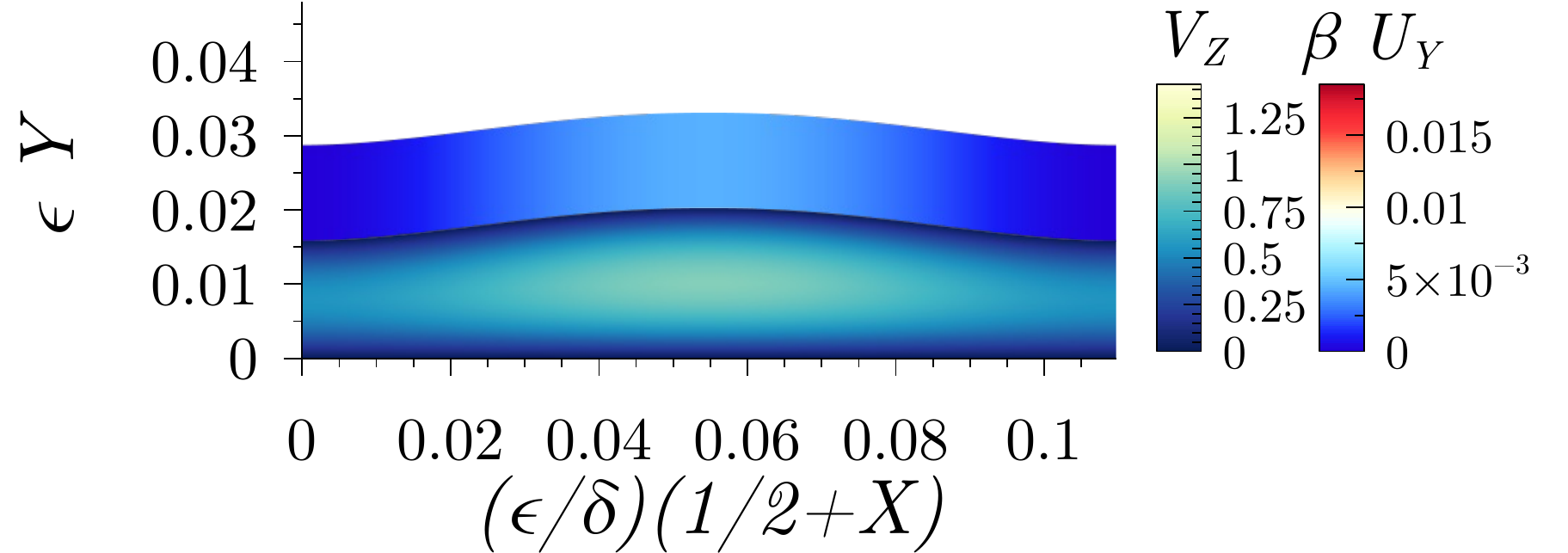} &   \includegraphics[width=0.36\textwidth]{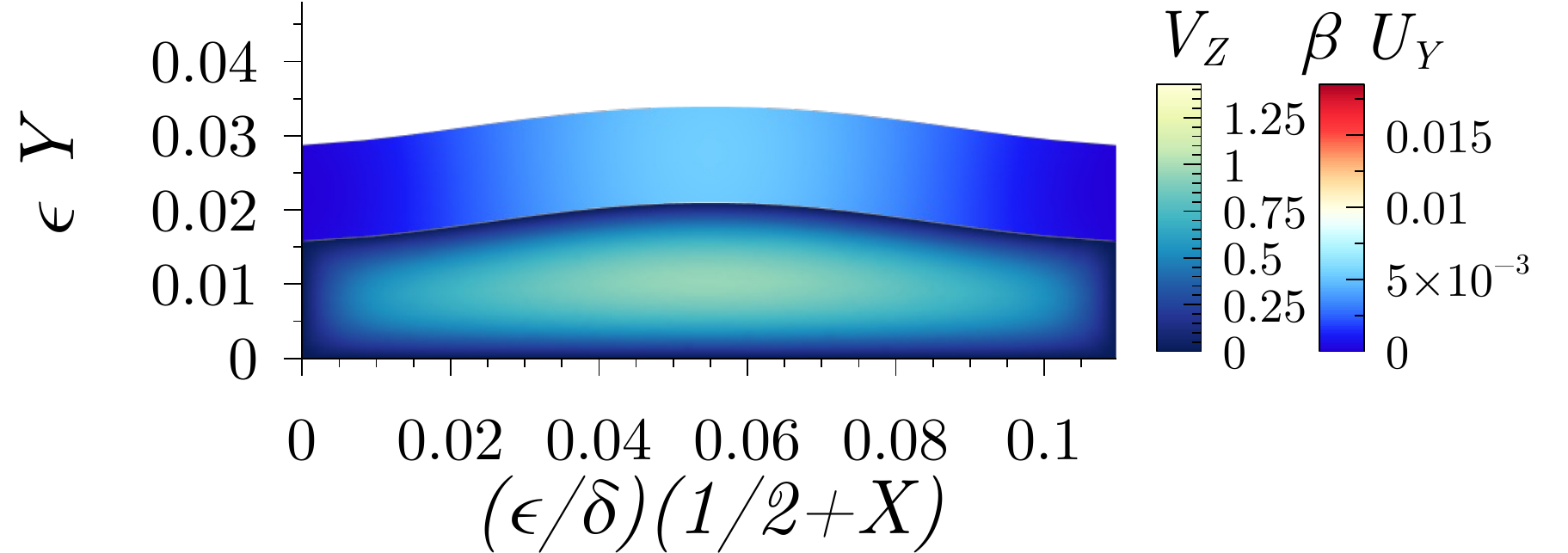} &  \includegraphics[width=0.36\textwidth]{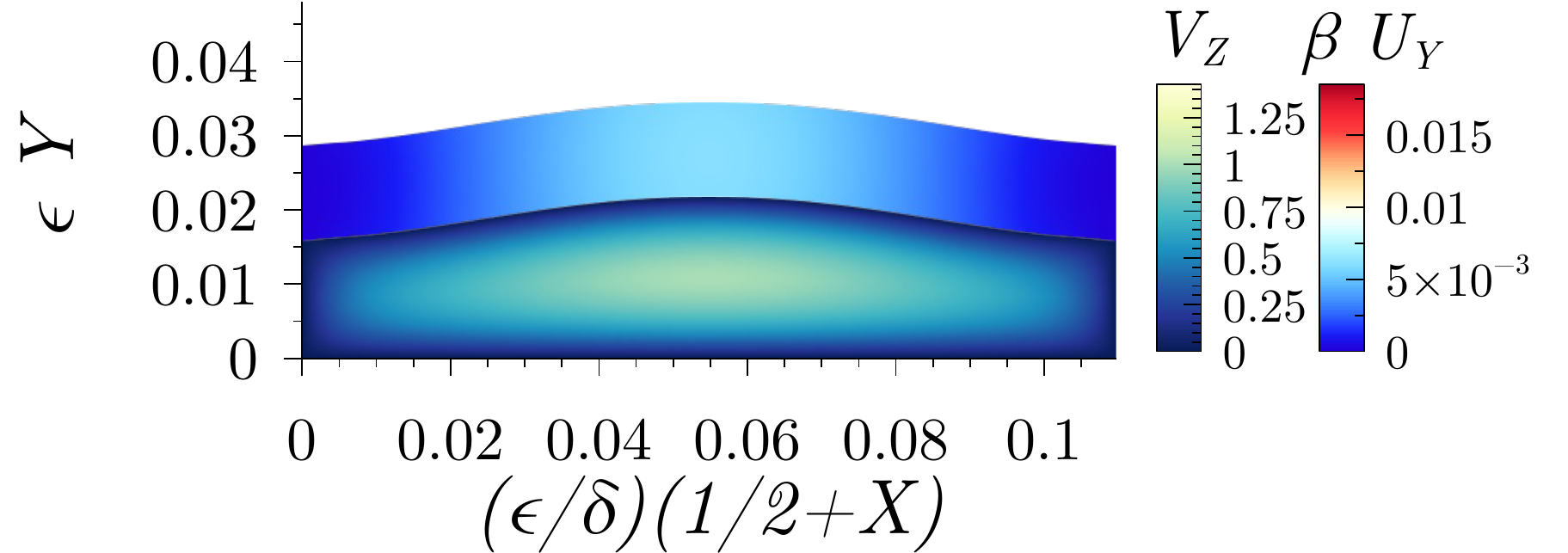} \\
  ($d$) & ($j$)  & ($p$)\\
 \includegraphics[width=0.36\textwidth]{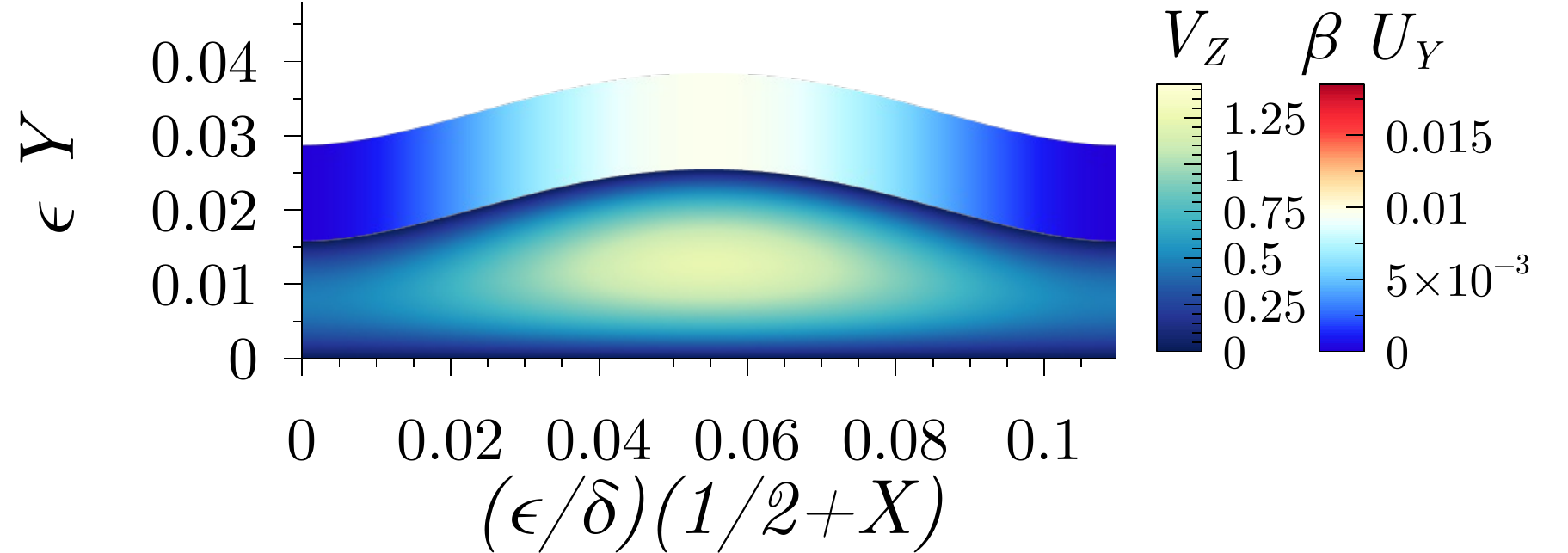} &   \includegraphics[width=0.36\textwidth]{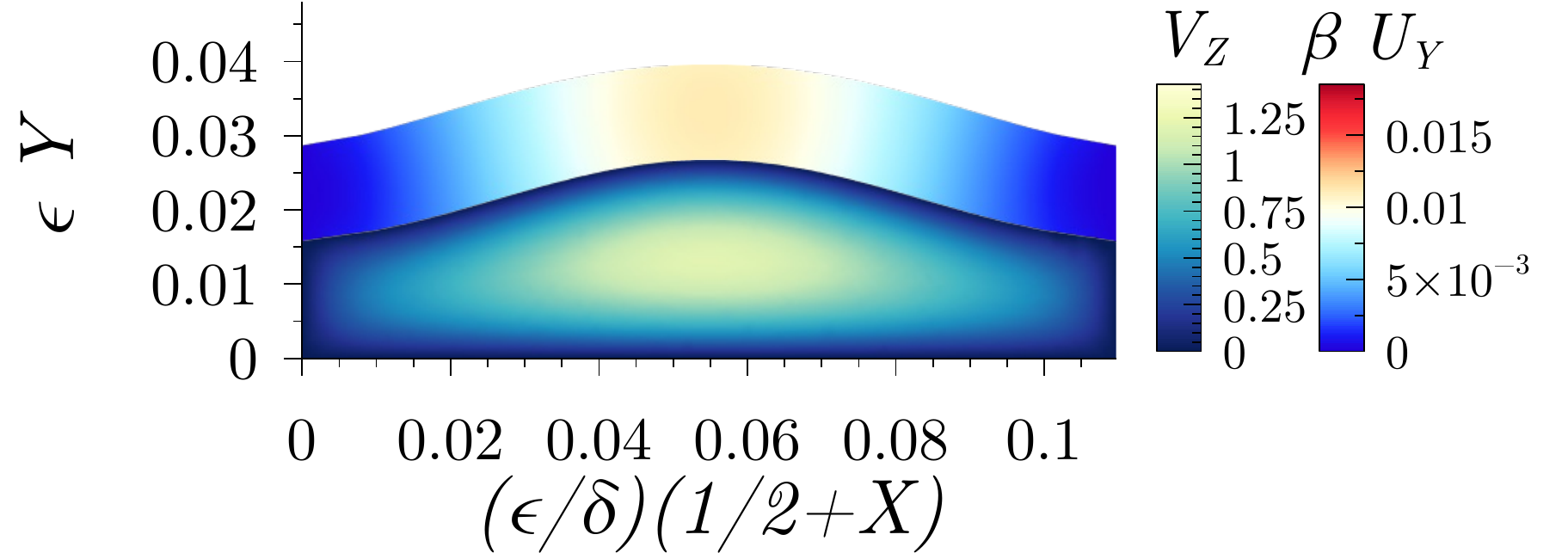} & \includegraphics[width=0.36\textwidth]{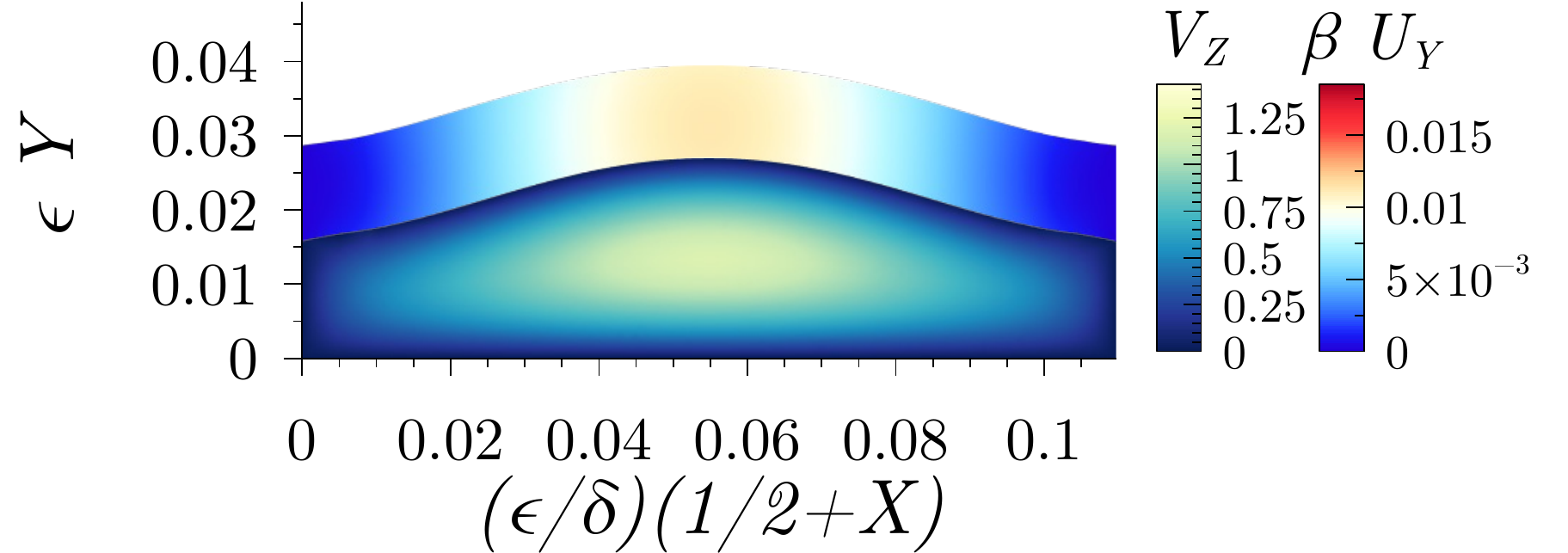}\\
   ($e$) & ($k$)  & ($q$)\\
 \includegraphics[width=0.36\textwidth]{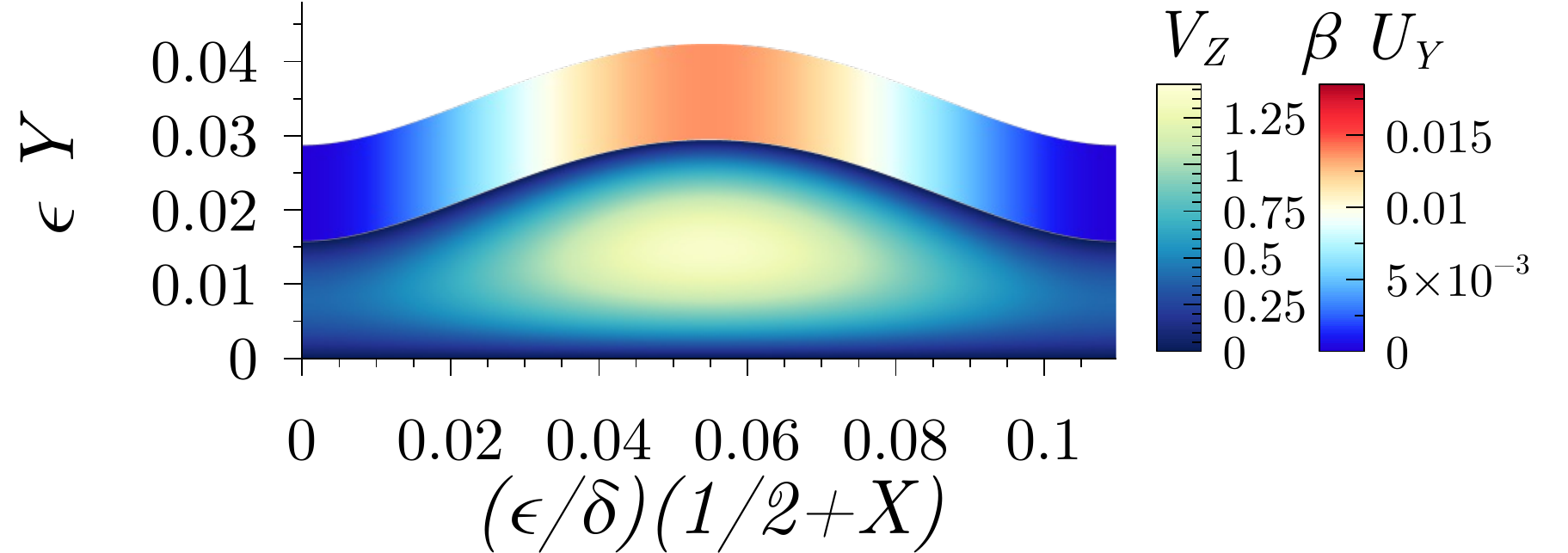} &   \includegraphics[width=0.36\textwidth]{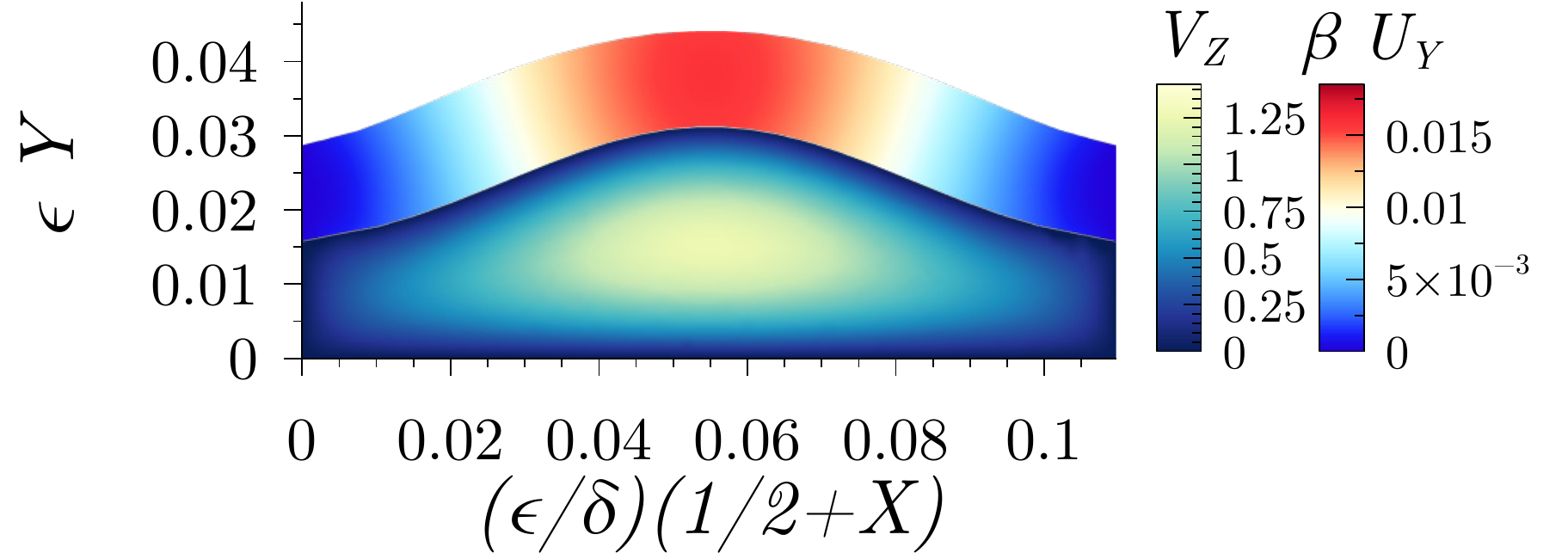} & \includegraphics[width=0.36\textwidth]{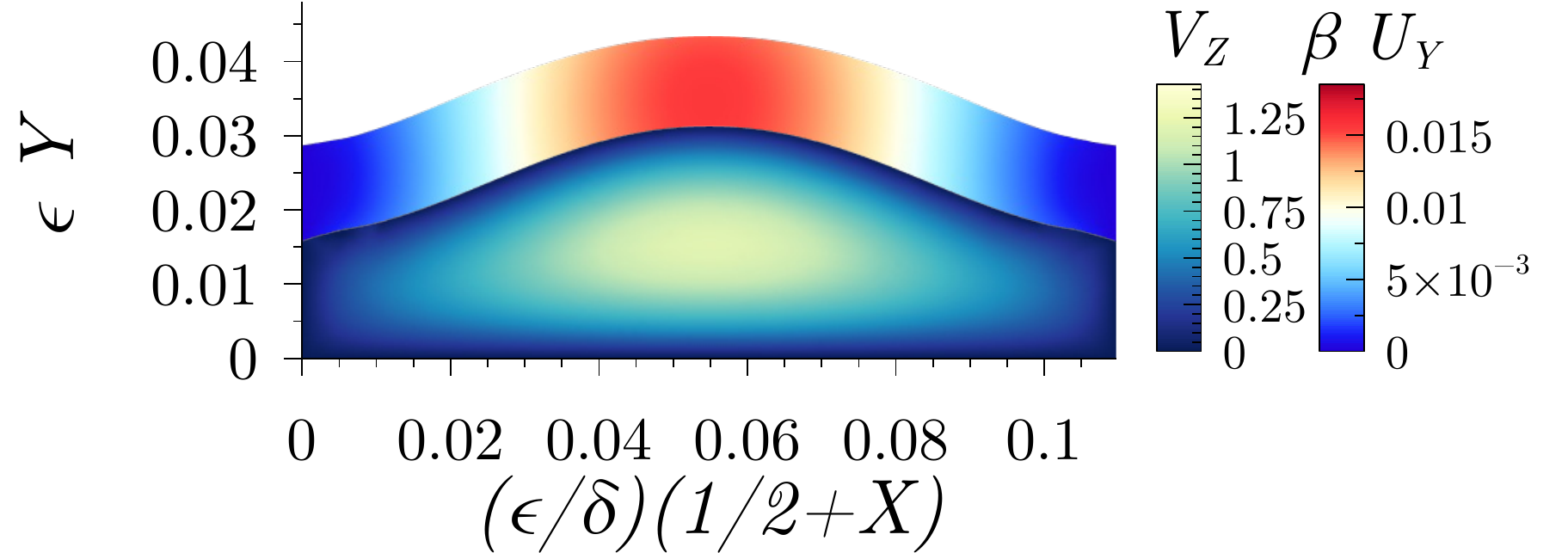}\\
   ($f$) & ($l$)  & ($r$)\\
 \includegraphics[width=0.36\textwidth]{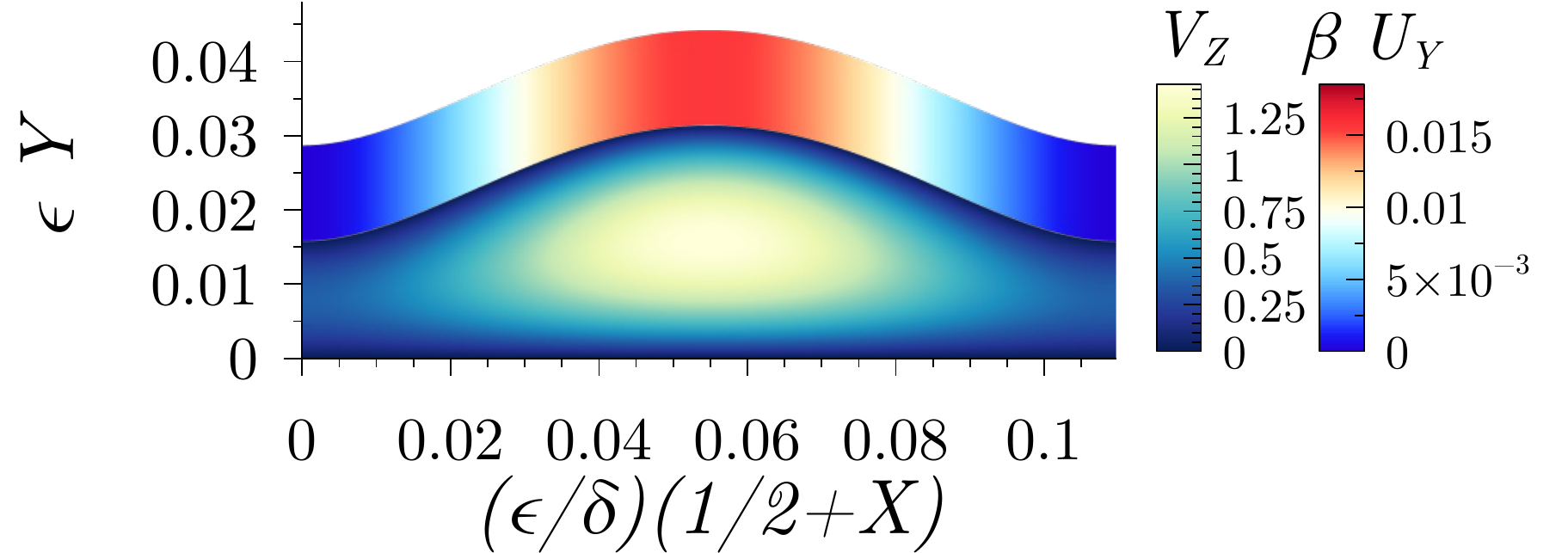} &   \includegraphics[width=0.36\textwidth]{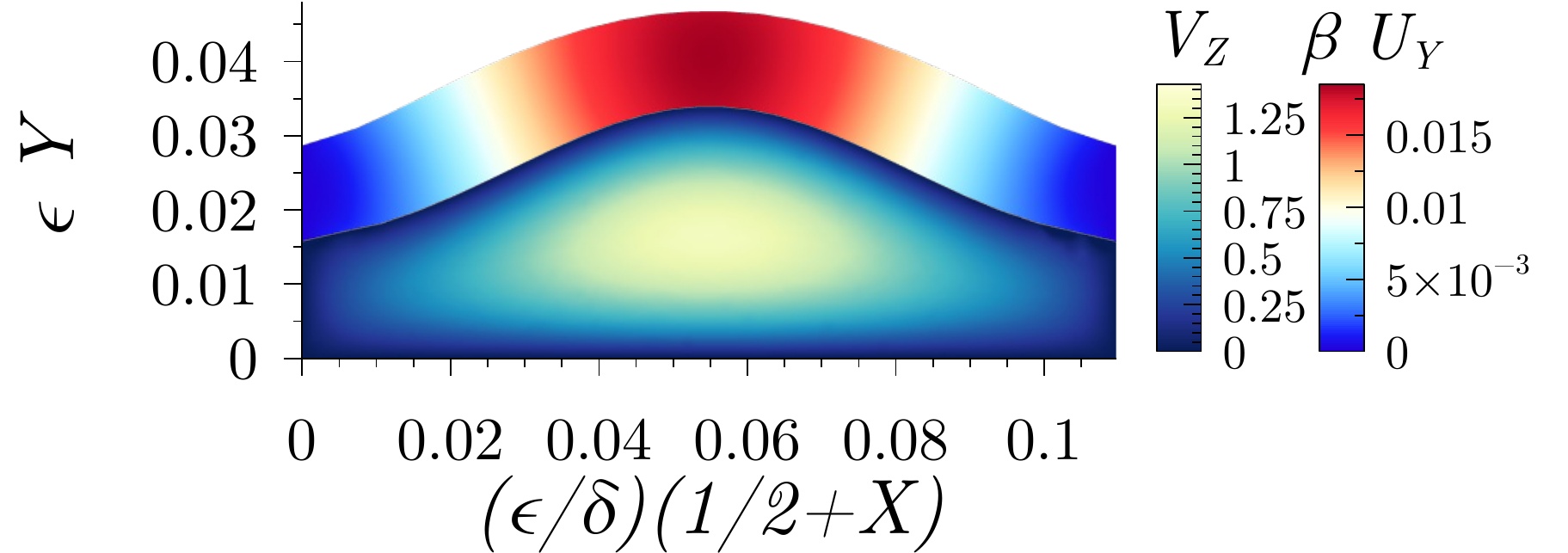} & \includegraphics[width=0.36\textwidth]{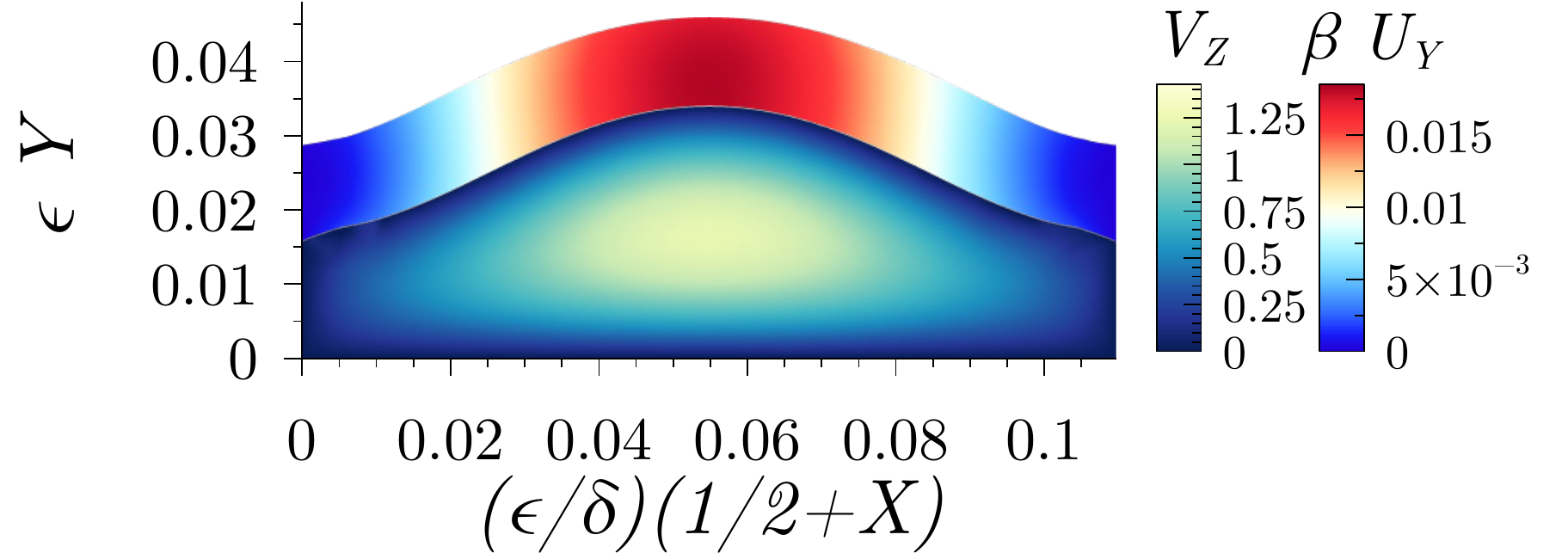}
\end{tabular}
\caption{\label{fig:fig9}(Colour online) Colourplot of longitudinal velocity $V_Z$ and vertical displacement $\beta U_Y$ in the $(X,Y)$ plane at a station $Z = 0.25$ and times $T = 3.5 \times 10^{-3}$, $1.8 \times 10^{-2}$, $5.3 \times 10^{-2}$, $1.4 \times 10^{-1}$, $2.8 \times 10^{-1}$ and $7.0 \times 10^{-1}$, in the column ($a$--$f$), extracted from the quasi-steady lubrication model~\eqref{eq:P_eq}--\eqref{eq:c}, in ($g$--$l$), extracted from 3D numerical simulations using the linear strain~\eqref{eq:linear_strain}, and in ($m$--$r$) using the nonlinear strain~\eqref{eq:nonlinear_strain}, for $\tilde{\beta} = 4.8$ under flow-rate-controlled conditions. The values of the parameters correspond to those of figure~\ref{fig:fig7}($b$).}
\end{figure}

In Case I, the agreement in the time evolution of $Q$, $P$ and $U_m$ is fairly good both in the pressure-driven and in the flow-rate-controlled configurations, as evidenced by figure~\ref{fig:fig6}. In particular, the self-similar solution $P(\zeta)$, $\varphi(\zeta)$ given by~\eqref{eq:self_similar} correctly describes the rescaled numerical solution, as shown by the insets in figure~\ref{fig:fig6}($a$). In this case, since $\epsilon Re/12, \epsilon Re/(4\tilde{\beta}),\tilde{\gamma} \ll 1$, the largest source of error is probably the fact that the lubrication approximation does not satisfy the no-slip condition for $\boldm{v}$ at the lateral walls, nor does it satisfy the clamped condition for $\boldm{u}$ at the inlet, which gives rise to an elastic boundary layer where the largest disagreement takes place, especially for $U_m$. However, as pointed out by~\cite{Christov2018}, its influence is confined to a region of length $O(\epsilon/\delta)\ll 1$. Additionally, the lubrication approximation is not able to capture the early-time oscillations experienced by $P$ and $U_m$ close to the inlet. These travelling waves are always present even when $\epsilon Re$ and $\tilde{\gamma}$ are exactly zero. Hence, a possible explanation might be that the derivatives $\partial^4_z u_y$ and $\partial^2_z \partial^2_x u_y$ are significant at early time since its characteristic length scale is $Z \sim 0.05-0.1$, which is of the same order as $\delta$ for the geometry considered in Cases I and II. Therefore, the $Z$-scale is initially small as the fluid spreads within the channel causing the oscillations, which eventually disappear as the fluid propagates and the $Z$-scale increases. An analogous phenomenon has been observed by~\cite{Lister2013}, where they found travelling-wave solutions for the peeling of an axisymetric elastic sheet.

In Case II, although the values of $\epsilon Re/12$, $\epsilon Re/(4\tilde{\beta})$, and $\tilde{\gamma}$ are not strictly small, and the lubrication hypotheses are not satisfied, the agreement between the 3D simulations and the quasi-steady approximation is better than might be expected, as shown by the results in figure~\ref{fig:fig7}. However, in this case, the amplitude and the dissipation time of the early-time oscillations of $P$ and $U_m$ are larger, and they also propagate downstream to larger values of $Z$. This behaviour breaks the self-similarity of $P$ and $Q$ in the pressure-controlled configuration.

Finally, to provide a better illustration of the agreement between the quasi-steady lubrication model and the 3D numerical simulations, we have also computed the time-dependent evolution of the longitudinal velocity $V_Z$ and the vertical displacement $\beta U_Y$ at a longitudinal station close to the inlet, $Z = 0.25$, under flow-rate-controlled conditions, and for the values of Case II in table~\ref{tab:table1}, i.e. the least favourable configuration, but for $\tilde{\beta} = 4.8$, in order to also test the validity of the linear-strain approximation~\eqref{eq:linear_strain}. These conditions correspond to an inlet flow rate of $q_0 = 429$ ml min$^{-1}$, whereas the maximum flow rate reported in the experiments of~\cite{Ozsun2013} is $q_0 = 50$ ml min$^{-1}$ (see figure~\ref{fig:fig2}). Figure~\ref{fig:fig9} displays six different time snapshots, the last one corresponding to the steady state, showing $V_Z$ and $\beta U_Y$ obtained from the quasi-steady lubrication model~\eqref{eq:P_eq}--\eqref{eq:c} in the column ($a$--$f$), from the 3D numerical simulations, using the linear strain~\eqref{eq:linear_strain}, in the column ($g$--$l$), and using the nonlinear strain, in the column ($m$--$r$). Taking into account that the hypotheses behind the lubrication model~\eqref{eq:P_eq}--\eqref{eq:c} are not strictly satisfied, the overall agreement between the quasi-steady model and the 3D simulations during the whole start-up transient is quite good. However, again, there are marked differences between both approaches, e.g. in the no-slip condition at the lateral bounding walls, which the lubrication model cannot fulfill, or the constant thickness and the unidirectional displacement of the top wall considered by the bending-dominated Kirchhoff--Love theory, which do not apply in this configuration where the top wall thickness is comparable to the channel width. Furthermore, the quasi-steady lubrication model slightly underestimates the axial velocity of the fluid and the vertical displacement. 

Comparing the 3D numerical results ($g$--$l$) and ($m$--$r$) for linear~\eqref{eq:linear_strain} and nonlinear~\eqref{eq:nonlinear_strain} strains, respectively, we find that in this case, even though $\tilde{\beta}$ is relatively large and the deflection is comparable to the thickness of the elastic wall, the nonlinear stretching of the wall has a negligible effect on its elastic response, and the linear-strain model is adequate.

\section{Conclusions}\label{sec:conclusions}

In this paper we have studied the start-up flow in a shallow rectangular microchannel with a deformable top wall, considering both pressure-controlled and flow-rate-controlled conditions. To that end, we have developed an unsteady lubrication model, where the top wall is modelled with the Kirchhoff--Love plate theory in the bending-dominated limit. To derive this simplified model we have first identified the characteristic scales and the dimensionless parameters governing the hydro-elastic problem showing, in particular, that the characteristic start-up time only depends on the geometry of the channel and on the solid and fluid properties, but not on the characteristic pressure and flow rate. When the solid and liquid inertia are negligible, the lubrication model is quasi-steady and reduces to a nonlinear diffusion equation for the fluid pressure field, whose only dimensionless parameter is the compliance parameter $\tilde{\beta}$ and that, under pressure-controlled conditions, admits a self-similar solution.

To check the validity of the hypotheses behind the lubrication model in the limit of negligible solid and liquid inertia, we have conducted 3D numerical simulations of the complete Navier and Navier-Stokes equations for the solid and for the fluid, respectively. In particular, as a basis for comparison, we have selected a microchannel whose geometry corresponds to the S4 experiment of~\cite{Ozsun2013}, and two working liquids. First, we have considered a silicon oil of 500 cSt dynamic viscosity, which fulfills the hypotheses, and we have found excellent agreement between the pressure, displacement, and flow rate predicted by the quasi-steady lubrication model, and those obtained from the 3D numerical simulations. In the second case we have considered water as working liquid, for which the liquid and solid inertia are not negligible, although its influence is moderate. In this case we have also obtained fair agreement. We have also derived a leading-order asymptotic solution in terms of a regular expansion in the compliance parameter $\tilde{\beta}$, which properly captures the transient dynamics of the microchannel when $\tilde{\beta} \ll 1$.

We have also computed the start-up time for several values of $\tilde{\beta}$, comparing the values predicted by our lubrication model with those obtained from the model of~\cite{Dendukuri2007} and from the 3D numerical simulations. In the two flow configurations considered herein, we have obtained good agreement between the simulations and our model, but not with the lubrication approximation of~\cite{Dendukuri2007}. The reason is that these authors assume that the pressure scales as $p_c \sim u_c E/w$, which is only valid if the elastic wall is large enough, but not when the thickness of the wall is smaller than, or of the same order as the channel width, where the appropriate scaling is $p_c \sim u_c B /w^4$.

There are many extensions of the present work that deserve further effort. First of all, the inertial corrections associated with finite values of $\tilde{\alpha}$ and $\tilde{\gamma}$ in the lubrication equations~\eqref{eq:momentumz_dimless}--\eqref{eq:bc_outlet_dimless} should be studied. In addition, other unsteady processes like pulsatile flows should be addressed. For intermediate and large wall thicknesses, the bending-dominated Kirchhoff--Love plate theory fails, and thus stretching and shear have to be included in the modelling, which could be addressed along the lines of~\cite{Shidhore2018} for steady flow. Finally, the case of a microchannel embedded in an elastic half-space should also be studied, since it occurs frequently in applications.

\begin{acknowledgements}
\section*{Acknowledgements} \label{app:acknow}

The authors are grateful to Javier Rivero-Rodr\'iguez and Benoit Scheid for key numerical advice, to Ivan C. Christov for pointing out a mistake in figure 1 of an earlier version of the manuscript, and to Ram\'on Zaera for helpful discussions. AM-C and AS thank the Spanish MINECO, Subdirecci\'on General de Gesti\'on de Ayudas a la Investigaci\'on, for its support through projects DPI2014-59292-C3-1-P and DPI2015-71901-REDT, and the Spanish MCIU-Agencia Estatal de Investigaci\'on through project DPI2017-88201-C3-3-R. These research projects have been partly financed through FEDER European funds. AM-C also acknowledges support from the Spanish MECD through the grant FPU16/02562 and to its associated program Ayudas a la Movilidad 2018 during his stay at the Complex Fluids Group in Princeton. HAS thanks the NSF for support via CMMI-166-1672 and through Princeton University's Material Research Science and Engineering Center DMR-1420541.
\end{acknowledgements}

-----------------------------

\begin{appendix}

\section{The small-compliance limit, $\tilde{\beta}\ll 1$}\label{app:asymp}
To study the limit $\tilde{\beta}\ll 1$, we expand the pressure field in $\tilde{\beta}$ as
\begin{equation}\label{eq:expansion}
P(Z,T;\tilde{\beta}) = P^{(0)}(Z,T) + \tilde{\beta} P^{(1)}(Z,T) + O(\tilde{\beta}^2).
\end{equation}
At leading order, using $F(\tilde{\beta} P) \approx 1$, the governing equation~\eqref{eq:P_eq} simplifies to the diffusion equation,
\begin{equation}
\partial_T P^{(0)} = \partial_Z^2 P^{(0)},
\end{equation}
with boundary and initial conditions
\begin{subequations}
\begin{gather}
\begin{rcases} 
      P^{(0)}(Z=0,T) = 1, & \text{pressure-controlled} \\
      -\partial_Z P^{(0)}(Z=0,T) = 1, & \text{flow-rate-controlled}
\end{rcases}, \\
P^{(0)}(Z=1,T) = 0, \quad 
P^{(0)}(Z,T=0) = 0.
\end{gather}
\end{subequations}
The solutions can be found using (for example) a Fourier expansion, and are given by
\begin{subequations}\label{eq:asymp_P0}
\begin{alignat}{2}
  \label{eq:pc_P0} 
P^{(0)} &= 1-Z - \sum_{k=1}^{\infty} \frac{2 \sin(k\pi Z) e^{-(k \pi)^2 T}}{k\pi} & & \quad \text{(pressure-controlled)},
\\
  \label{eq:fr_P0}
P^{(0)} &= 1-Z - \sum_{k=1}^{\infty} \frac{8 \cos[(2k-1)\pi Z/2] e^{-(2k-1)^2\pi^2 T/4}}{(2k-1)^2\pi^2} && \quad \text{(flow-rate-controlled)}.
\end{alignat}
\end{subequations}
The first-order correction $P^{(1)}$ satisfies the equation 
\begin{equation}
 \partial_T P^{(1)} - \partial_Z^2 P^{(1)} = \partial_Z( P^{(0)} \partial_Z P^{(0)}), \label{eq:beta_eqP}
\end{equation}
with suitable boundary conditions, and can in principle be calculated in the same way, but the leading-order result is sufficient to verify our numerical results for $\tilde\beta \ll 1$.

\begin{figure*}
\begin{tabular}{cc}
($a$) & ($b$)\\
\includegraphics[width=0.5\textwidth]{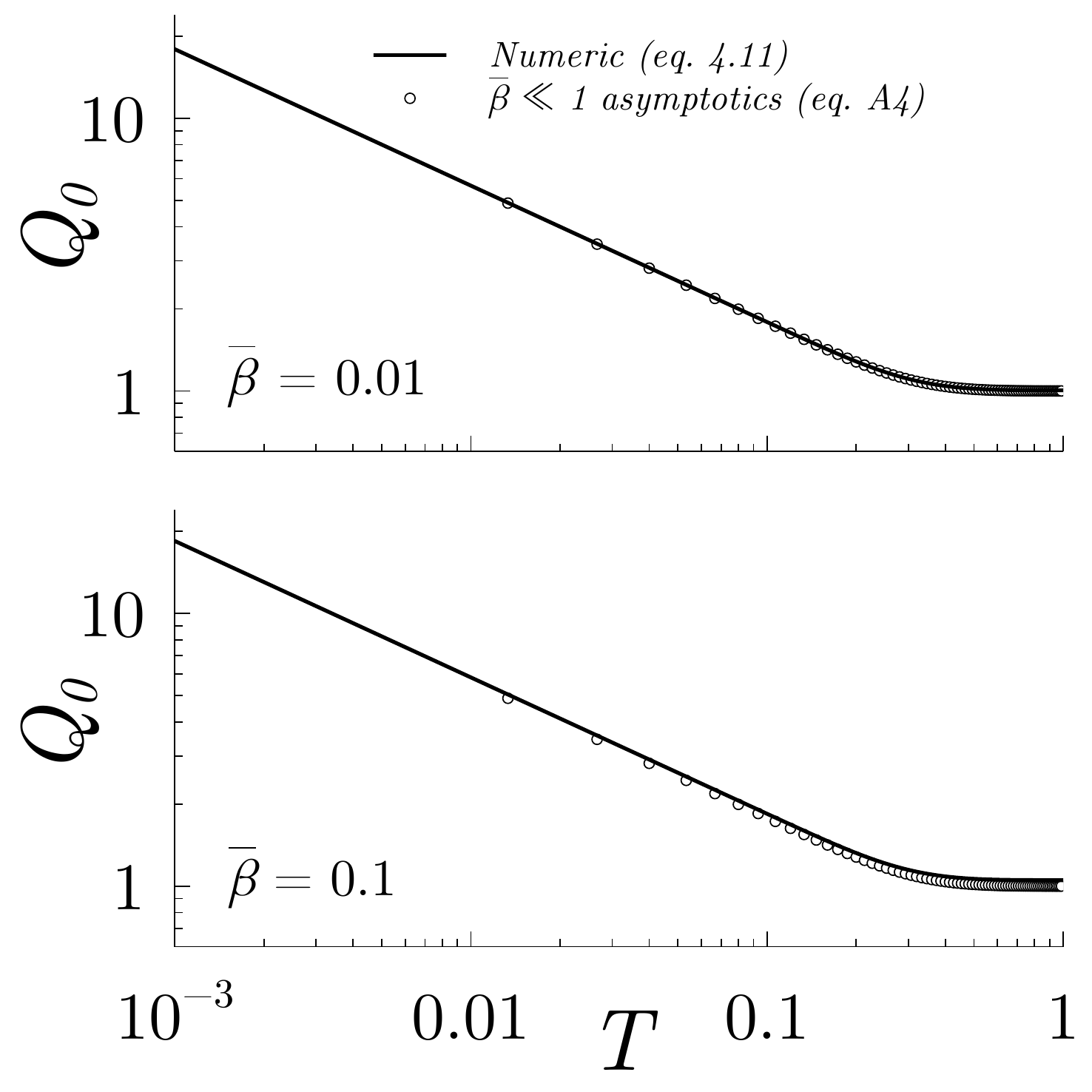} &
\includegraphics[width=0.5\textwidth]{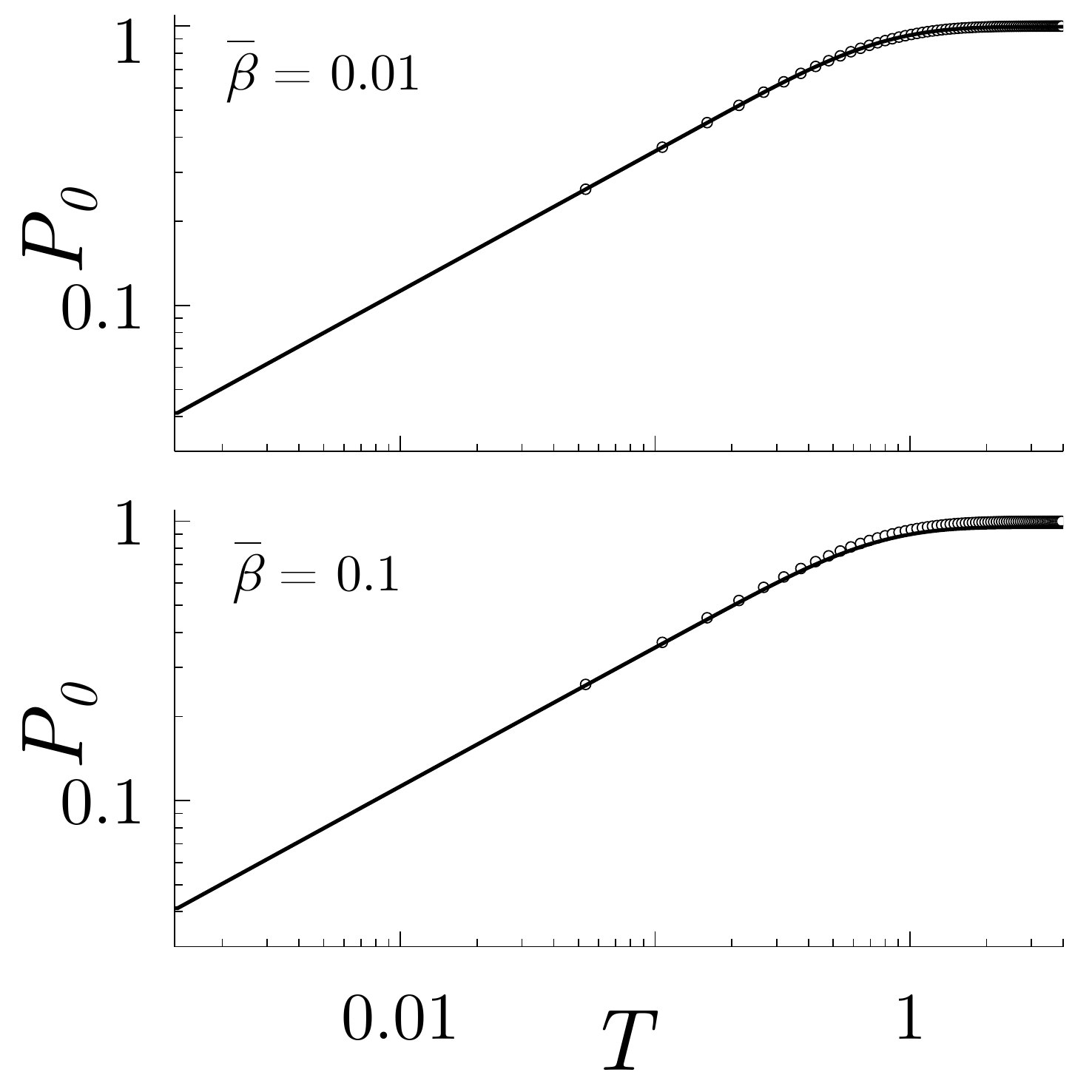}
\end{tabular}
\caption{Comparison of the inlet flow rate $Q_0(T)$ in a pressure-controlled configuration ($a$) and the inlet overpressure $P_0(T)$ in a flow-rate-controlled configuration ($b$), between numerical computations of~\eqref{eq:P_eq}--\eqref{eq:c} and its corresponding asymptotic solutions at leading order,~\eqref{eq:pc_P0}, and~\eqref{eq:fr_P0}\label{fig:fig8}.}
\end{figure*}

Figure~\ref{fig:fig8} shows a comparison of the inlet flow rate $Q_0(T)$ in a pressure-controlled configuration ($P_0 = 1$) and the inlet overpressure $P_0(T)$ in a flow-rate-controlled configuration ($Q_0 = 1$), between numerical computations of~\eqref{eq:P_eq}--\eqref{eq:c} and the corresponding leading-order solutions~\eqref{eq:pc_P0} and \eqref{eq:fr_P0}, respectively. These solutions have been used to obtain the start-up times $T_d$ in the limit $\tilde{\beta}\ll 1$, shown in figure~\ref{fig:fig3}($c$,$d$) (dashed line). Note that the leading-order asymptotic solutions work reasonably well when $\tilde{\beta} \lesssim 0.1$.

\section{Numerical implementation}\label{app:numerics}
In this appendix we describe the numerical techniques used to implement the system~\eqref{eq:continuity_momentum}--\eqref{eq:stress_free}. All the equations are written in weak form by means of the corresponding integral scalar product, defined in terms of test functions for the pressure, velocity, and displacement fields, i.e. $\tilde{p}$, $\tilde{\boldm{v}}$, and $\tilde{\boldm{u}}$, respectively. By using Green identities we finally obtain an integral bilinear system of equations for the set of variables and their corresponding test functions. Equation~\eqref{eq:continuity_momentum} reads in weak form
\begin{equation}
\int_{{\Omega}_f} [ \tilde{p} \, \bnabla \bcdot \boldm{v} + \rho ( \partial_t \boldm{v} + \boldm{v} \bcdot \bnabla \boldm{v}) \bcdot \tilde{\boldm{v}} +\mathsfbi{T} \boldm{:} \bnabla \tilde{\boldm{v}}] \, \text{d} \Omega_f - \sum_i \int_{\Gamma_f^i} (\mathsfbi{T} \bcdot \boldm{n}_f^i) \bcdot \tilde{\boldm{v}} \, \text{d} \Gamma_f = 0
\end{equation}
where $\Omega_f$ is the deformable 3D fluid domain, and $\Gamma_f^i$ are the boundary surfaces with their corresponding normal vectors $\boldm{n}_f^i$. The flux integral is set to zero at the fluid-solid interface, since the continuity of stress is imposed in the weak-form Navier equation. At the lateral and lower walls the no-slip condition for the velocity field~\eqref{eq:bc1} is taken into account by imposing $\boldm{\tilde{v}} = \boldm{0}$ at the corresponding boundaries. At the outlet we impose $\tilde{p} = 0$ as boundary condition for the pressure field, and also the stress-free boundary condition~\eqref{eq:bc_outlet}. Furthermore, at the inlet, in a pressure-controlled situation we impose a non-homogeneous Dirichlet boundary condition for the pressure and a stress-free boundary condition~\eqref{eq:ic_bc_1}. On the other hand, when the microchannel is flow-rate controlled we impose a normal velocity $v_0(x,y)$ corresponding to flow in a rigid channel with the desired flow rate, i.e.
\begin{equation}
(\partial_x^2 + \partial_y^2)v_0 = \text{constant}, \quad v = 0 \text{ at } \begin{cases} x = \pm w/2, \\ y = 0,\ h_0, \end{cases} \quad \int_{-w/2}^{w/2} \int_0^{h_0} v_0\,\text{d}y \, \text{d}x = q_0,
\end{equation}
and homogeneous Dirichlet boundary conditions for the tangential velocity~\eqref{eq:ic_bc_2}.

The Navier equation~\eqref{eq:navier} reads in weak form
\begin{equation}
\int_{\Omega_s} [  \rho_s ( \partial_t^2 \boldm{u}) \bcdot \tilde{\boldm{u}} + \boldm{\sigma} \boldm{:} \bnabla \tilde{\boldm{u}}] \text{d} \Omega_s -\sum_i \int_{\Gamma_s^i} (\boldm{\sigma} \bcdot \boldm{n}_s^i) \bcdot \tilde{\boldm{u}} \, \text{d} \Gamma_s^i = 0,
\end{equation}
where $\Omega_s$ is the 3D solid domain, and $\Gamma_s^i$ the boundaries with $\boldm{n}_s^i$ the corresponding normal vector. The four clamping conditions~\eqref{eq:clamped_3D} are imposed as $\tilde{\boldm{u}} = \boldm{0}$ in the flux integral, whereas the continuity of stress~\eqref{eq:continuity_stress} at the contact interface and the stress-free condition~\eqref{eq:stress_free} at the outer wall are imposed as natural boundary conditions, which read, respectively
\begin{equation}
\int_{\Gamma_s^{\text{cont}}} (\boldm{\sigma} \bcdot \boldm{n}) \bcdot \tilde{\boldm{u}} \, \text{d} \Gamma_s^{\text{cont}}  = -\int_{\Gamma_s^{\text{cont}}} (\mathsfbi{T} \bcdot \boldm{n}) \bcdot \tilde{\boldm{u}} \, \text{d} \Gamma_s^{\text{cont}} \quad \text{at} \quad y = h(x,z,t),
\end{equation}
\begin{equation}
\int_{\Gamma_s^{\text{ext}}} (\boldm{\sigma} \bcdot \boldm{n}_{\text{ext}}) \bcdot \tilde{\boldm{u}} \, \text{d} \Gamma_s^{\text{ext}} = 0 \quad \text{at} \quad y = h(x,z,t) + d(x,z,t).
\end{equation}
Moreover, the continuity of velocity~\eqref{eq:bc2} is imposed as a weak constraint. 

The equations are discretised using Taylor-Hood tetrahedral elements for the pressure and velocity fields, and second-order Lagrange elements for the displacement field, which ensures numerical stability. To account for the deformation of the domain we use the ALE method implemented in COMSOL Multiphysics, where the mesh elements in the solid domain $\Omega_s$ move with an imposed displacement given by $\boldm{u}$, whereas in the fluid domain $\Omega_f$ they move according to the Laplace equation for the change of variable between the material and the spatial frames~\citep{RiveroScheid2018, Rivero2019}. As for the time-stepping, we employ a 4th-order variable-step BDF method, or an implicit generalised-alpha method when the inertia of the solid becomes relevant (Case II). The relative tolerance of the nonlinear method is always set to $10^{-5}$.
\end{appendix}

\bibliographystyle{jfm}
\bibliography{biblio}

\end{document}